\documentclass{pasj01}
\usepackage{bm}

\usepackage[switch,mathlines]{lineno}   
\usepackage{color}
\usepackage{natbib}
\bibpunct[:]{(}{)}{;}{a}{}{,}

\begin{document}

\title{ 
Radiation and outflow properties of super-Eddington accretion flows around various mass classes of black holes:                             
Dependence on the accretion rates   
}

\author{Shogo \textsc{YOSHIOKA}\altaffilmark{1}}
\altaffiltext{1}{Department of Astronomy, Graduate School of Science, Kyoto University, Kitashirakawa-Oiwake-cho, Sakyo-ku, Kyoto 606-8502, Japan}
\email{yoshioka@kusastro.kyoto-u.ac.jp}

\author{Shin \textsc{Mineshige}\altaffilmark{1}}

\author{Ken \textsc{Ohsuga}\altaffilmark{2}}
\altaffiltext{2}{Center for Computational Sciences, University of Tsukuba, Ten-nodai, 1-1-1 Tsukuba, Ibaraki 305-8577, Japan}

\author{Tomohisa \textsc{Kawashima}\altaffilmark{3}}
\altaffiltext{3}{Institute for Cosmic Ray Research, The University of Tokyo, 5-1-5 Kashiwanoha, Kashiwa, Chiba 277-8582, Japan}

\author{Takaaki \textsc{Kitaki}}


\KeyWords{accretion, accretion disks --- radiation dynamics --- stars: black holes}

\maketitle

\begin{abstract}
  We perform axisymmetric two-dimensional radiation-hydrodynamic simulations of super-Eddington accretion flow and outflow 
  around black holes to examine the properties of radiation and outflow 
  as functions of the black hole mass and the accretion rate onto the black hole ($\dot M_{\rm BH}$).
  We find that the $\dot{m}_{\rm BH} (\equiv \dot{M}_{\rm BH}c^2 /L_{\rm Edd})$ dependence of $L_{\rm rad}/L_{\rm Edd}$ and $L_{\rm mech}/L_{\rm Edd}$ found for stellar-mass black hole can apply to the high mass cases, 
  where $L_{\rm rad}$ is the radiation luminosity, $L_{\rm mech}$ is the mechanical luminosity, $c$ is the speed of light, and $L_{\rm Edd}$ is the Eddington luminosity.
  Such universalities can appear in the regime, in which electron scattering opacity dominates over absorption opacity.
  Further, the normalized isotropic mechanical luminosity
  $L_{\rm mech}^{\rm ISO}/L_{\rm Edd}$ (evaluated by normalized density and velocity at $\theta=10^\circ$) 
  exhibits a broken power-law relationship with ${\dot m}_{\rm BH}$; 
  $L_{\rm mech}^{\rm  ISO}/ L_{\rm Edd} \propto{\dot m}_{\rm BH}^{2.7}$ (or $\propto {\dot m}_{\rm BH}^{0.7}$) 
  below (above) ${\dot m}_{\rm BH}\sim 400$.
  This is because the radial velocity stays nearly constant (or even decreases) below (above) the break with increase of $\dot m_{\rm BH}$.
  We also find that the luminosity ratio is $L_{\rm mech}/L_{\rm rad}^{\rm  ISO} \sim$ 0.05 
  at ${\dot m}_{\rm BH} \sim 100$, which is roughly consistent with the observations of NLS1, 1H 0323+103.

\end{abstract}


\section{Introduction}
\label{sec-introduction}
It has been long believed that the Eddington luminosity is a classical limit to the luminosities of any accreting objects. 
However, we now know that this is no longer the case both from the observational and theoretical grounds. 
In fact, there are growing observational evidences supporting the existence of super-Eddington accretion in several distinct classes of objects.

Good candidates for super-Eddington accretors are Ultra-Luminous X-ray sources (ULXs), bright X-ray compact objects with X-ray luminosities of $10^{39} - 10^{41}~ [\rm erg~ s^{-1} ]$ \citep[see recent review by][]{kaaret2017}.
ULX are located at a off-center regions; that is, they are not active galactic nucleus \citep{long1996, fabbiano1989, soria2007, kaaret2017}. 
Their central engines are still under discussion; 
romising models include (1) super-Eddington accretion onto a stellar mass black hole \citep{watarai2001,king2001, gladstone2009, kawashima2012, sutton2013, motch2014, middleton2015, kitaki2017}, 
(2) super-Eddington accretion onto a neutron star \citep{Bsk1976, Bachetti2014, furst2016, kawashima2016, israel2017, carpano2018, mushtukov2018, inoue2023}, 
and (3) sub-Eddington accretion onto an intermediate-mass black hole\citep{makishima2000, miller2004, strohmayer2009, miyawaki2009}.

Other super-Eddington accretors are found in narrow-line Seyfert 1 galaxies (NLS1s) such as 1H 0323+342.  
NLS1s tend to have higher Eddington ratios than the Seyfert with similar luminosities, thus super-Eddington accretion being more feasible in the former \citep{wang1999, mineshige2000}. 
\citet{jin2022}, for example, analyzed RX J0134.2-4258 (NLS1) using the multi-wavelength spectrum and estimated the accretion rate to be $\sim 20-26~L_{\rm Edd}/c^2$ in the outer disk (for the black hole mass $M_{\rm BH}/M_\odot \sim 10^7$).
Although a majority of NLS1s are radio quiet, $\sim 7\%$ of them are radio-bright objects \citep{zhou2006, yuan2008}.
It is observationally suggested that some NLS1s have a fast outflow component with velocities greater than $0.1c$ called Ultra Fast Outflow (UFO) \citep[e.g.,][]{cappi2009, tombesi2014}.
NLS1 is expected to influence the evolution of its host galaxies through the process called AGN feedback 
because the outflow with high velocity, including UFOs, carries a large amount of kinetic energy \citep{nardini2015, tombesi2015}.
It is also known that some radio-loud NLS1s have relativistic jets and emit high energy radiation such as $\gamma$ ray emission \citep{lister2016, DAmmand2019}.
In many aspects, NLS1 exhibits interesting observational features.

To understand the nature of super-Eddington accretors, 
it is essential to solve the interaction between the radiation and the gas; that is, the radiation hydrodynamics (RHD) simulations are necessary \citep{eggum1988, fujita1998, ohsuga2005, ogawa2017, takeo2018, kitaki2018}.
Such RHD simulations have been extensively performed in these days, followed by general relativistic radiation magnetohydrodynamics (GR-RMHD) simulations \citep[e.g.,][]{mckinney2014, mckinney2015, sadowski2014, sadowski2015, sadowski2016, takahashi2016, utsumi2022}.
We point out that these studies apply a small Keplerian radius in a small computational domain.
A small Keplerian radius indicates a small angular momentum of the injected gas.
It is thus difficult to investigate the case of the NLS1, in which the injected materials, presumably supplied from the torus, seem to have relatively large angular momenta \citep{takahashi2010, jin2017}.

To investigate the large-scale accretion disk and outflow around a 10 $M_\odot$ black hole, Kitaki et al. (\citeyear{kitaki2021}, hereafter K21) performed two-dimensional (2D) axisymmetric RHD simulations, 
assuming (1) a large Keplerian radius, $r_{\rm K} = 2430~r_{\rm S}$, 
and adopting (2) a large simulation box of a size of $r_{\rm out} = 3000~ r_{\rm S}$ so that they could elucidate the disk-outflow structure over a wide region. 
Yoshioka et al. (\citeyear{yoshioka2022}, hereafter Paper I) performed the same type of 2D-RHD simulations but for a variety of black hole accretion rates $\dot{M}_{\rm BH}$.
In Paper I we reported the following results, some of which were unexpected before our study.
\begin{enumerate}
  \item The mechanical luminosity grows more rapidly than the radiation luminosity with an increase of $\dot{M}_{\rm BH}$.
  \item When seen from a nearly face-on direction, the isotropic mechanical luminosity grows in proportion to $\dot{M}_{\rm BH}^{2.7}$, while the total mechanical luminosity is proportional to $\dot{M}_{\rm BH}^{1.7}$. 
\end{enumerate}

These simulations were restricted to the cases of a $10 ~M_\odot$ black hole, however.
As mentioned above, there are many super-Eddington accretors with massive black holes (e.g., NLS1).
It is nontrivial whether the accretion-rate dependence of luminosity shown in Paper I holds for massive black holes.
We, therefore, expand the parameter space to obtain a more general view of super-Eddington outflow. 
This is the primary objective of this study.

Another objective is to examine if the fitting formulae for luminosities 
found in Paper I hold in a wider range of mass accretion rates. The plan of
this paper is as follows:
We explain calculated models and numerical methods in section 2, 
and present our results of the accretion rate dependence of luminosity for various mass range in section 3. 
We then give discussion in section 4.
The main issues to be discussed are 
whether the normalized accretion rate dependence of luminosity is sensitive to black hole mass, 
variations in the luminosities and outflow structure with increasing normalized accretion rate,
and the connection with the observations of NLS1.
Throughout this paper, we use the normalized black hole mass and accretion rates defined as follows:
\begin{eqnarray}
  &m_{\rm BH}& \equiv \frac{M_{\rm BH}}{M_\odot}, \\ 
  &\dot{m}_{\rm BH}& \equiv \frac{\dot{M}_{\rm BH}}{L_{\rm Edd}/c^2}.
\end{eqnarray}

\section{Calculated Models and Numerical Methods}
\subsection{Radiation hydrodynamics Simulations}
\label{sec-rhd}
In the present study, we consider super-Eddington accretion flow and associated outflow around
black hole with mass of $m_{\rm BH} = 10, 10^4,$ and $10^7$.
We inject mass with a certain amount of angular momentum from the outer simulation boundary at a constant accretion rate $\dot{M}_{\rm input}$
(more quantitative description will be given later).
For calculating radiation flux and pressure tensors, we adopt the
flux-limited diffusion approximation \citep[FLD;][]{levermore1981, turner2001}.
Since we do not solve the magnetic fields in the present simulation, 
we adopt the $\alpha$ viscosity prescription \citep{shakura1973}.
General relativistic effects are taken into account by employing the pseudo-Newtonian potential \citep{paczynsky1980}.

Basic equations and numerical methods are the same as those in K21 \citep[see also][]{ohsuga2005, kawashima2009}.
We solve the axisymmetric two-dimensional radiation hydrodynamics equations     
in the spherical coordinates $(x,y,z)=(r\sin\theta\cos\phi, r\sin\theta\sin\phi, r\cos\theta)$,
where the azimuthal angle $\phi$ is set to be constant and the $z$-axis coincides with the rotation-axis.
We put a black hole at the origin. 
In this paper, we distinguish $r$, radius in the spherical coordinates, and $R\equiv\sqrt{x^{2}+y^{2}}$, radius in the cylindrical coordinates. 

The basic equations are explicitly written as follows:
The continuity equation is
\begin{eqnarray}
  \frac{\partial \rho}{\partial t}+\nabla\cdot\left(\rho\bm{v}\right)=0,
\end{eqnarray}
where $\rho$ and $\bm{v}=(v_{r},v_{\theta},v_{\phi})$ 
is the gas mass density and the velocity of gas, respectively. 

The equations of motion are
\begin{eqnarray}
\frac{\partial(\rho v_{r})}{\partial t}+\nabla\cdot\left(\rho v_{r}\bm{v}\right)
= -\frac{\partial p}{\partial r}
+\rho\left[\frac{v_{\theta}^2}{r}+\frac{v_{\phi}^2}{r}
-\frac{GM_{\rm BH}}{(r-r_{\rm S})^2}\right] \nonumber\\
+\frac{\chi}{c}F_{0}^{r},  \label{eom_r}\\
\frac{\partial (\rho rv_{\theta})}{\partial t}+\nabla\cdot\left(\rho rv_{\theta}\bm{v}\right)
= -\frac{\partial p}{\partial \theta}+\rho v_{\phi}^{2}\cot\theta 
+r\frac{\chi}{c}F_{0}^{\theta}, \label{eom_th}\\
\frac{\partial(\rho r\sin\theta ~ v_{\phi})}{\partial t}+\nabla\cdot(\rho r\sin\theta ~ v_{\phi}\bm{v})
= \frac{1}{r^{2}}\frac{\partial}{\partial r}\left(r^{3}\sin\theta ~ t_{r\phi}\right),
\end{eqnarray}
where $p$ is the gas pressure,
$\chi=\kappa_{\rm abs}+\rho \sigma_{\rm T}/m_{\rm p}$ is the total opacity,
\citep[with $\kappa_{\rm abs}$ being free-free and free-bound absorption opacity and
$\sigma_{\rm T}$ being the cross-section of Thomson scattering, see][]{rybicki1991},
$m_{\rm p}$ is the proton mass, and
$\bm{F}_{0}=(F_{0}^r,F_{0}^\theta,F_{0}^{\phi})$ is the radiative flux in the comoving frame. 
In the present study, we use a simplified expression of the opacity (equation 6). 
If we would use a sophisticated opacity table, it may affect the quantitative results. 
This will be discussed at the end of section 4.1.
In the viscous-shear stress, only the $r$-$\phi$ component is assumed to be nonzero and is prescribed as
\begin{eqnarray}
  t_{r\phi}&=&\eta r \frac{\partial }{\partial r}\left(\frac{v_{\phi}}{r} \right),
\end{eqnarray}
with the dynamical viscous coefficient being
\begin{eqnarray}
  \eta&=&\alpha \frac{p+\lambda E_{0}}{\Omega_{\rm K}},
  \label{dvis}
\end{eqnarray}
where $\alpha=0.1$ is the viscosity parameter,
$\Omega_{\rm K}$ is the Keplerian angular speed,
$E_{0}$ is the radiation energy density in the comoving frame,
and $\lambda$ represents the FLD approximation.
We adopted the functional form, equation ({\ref{dvis}}), assuming that 
$\eta$ is proportional to the total pressure in the optically thick limit (since then we have $\lambda=1/3$ ), 
and that $\eta$ is proportional to the gas pressure in the optically thin limit (since then we find $\lambda = 0$) 
so that the adopted prescription should agree with that employed in the standard disk theory. 
Note that we have adopted the same prescription in the previous simulation studies \citep[e.g.,][]{ohsuga2005, kawashima2009, kitaki2021}.
In practice, local radiation MHD simulations demonstrate that 
the magnitude of the effective alpha viscosity is proportional to the total pressure 
(i.e., radiation plus gas pressure) rather than the gas pressure only in the limit of optically thick, 
radiation pressure dominant disks \citep[e.g.,][]{turner2004}.

The energy equation for gas is
\begin{eqnarray}
  \frac{\partial e}{\partial t}+\nabla\cdot\left(e\bm{v}\right)&=&-p\nabla\cdot\bm{v}-4\pi\kappa B+c\kappa E_{0}\nonumber\\
  &&+\Phi_{\rm vis}-\Gamma_{\rm Comp},
\end{eqnarray}
while the energy equation for radiation is
\begin{eqnarray}
  \frac{\partial E_{0}}{\partial t}+\nabla\cdot\left(E_{0}\bm{v}\right)&=&-\nabla\cdot\bm{F}_{0}-\nabla\bm{v}:\bm{{\rm P}}_{0}+4\pi\kappa B-c\kappa E_{0}\nonumber\\
  &&+\Gamma_{\rm Comp}.
\end{eqnarray}
Here $e$ is the internal energy density,
which is linked to the gas pressure through the ideal gas equation of state,
$p=(\gamma -1)e=\rho k_{\rm B}T_{\rm gas}/(\mu m_{\rm p})$
(with $\gamma=5/3$ being the specific heat ratio,
$k_{\rm B}$ being the Boltzmann constant,
$\mu=0.5$ being the mean molecular weight (we assume pure hydrogen plasmas),
and $T_{\rm gas}$ being the gas temperature, respectively),
$B=\sigma_{\rm SB}T_{\rm gas}^{4}/\pi$ is the blackbody intensity
(with $\sigma_{\rm SB}$ being the Stefan--Boltzmann constant),
$\bm{{\rm P}}_{0}$ is the radiation pressure tensor in the comoving frame, and
$\Phi_{\rm vis}$ is the viscous dissipative function;
\begin{eqnarray}
  \Phi_{\rm vis}=\eta\left[r\frac{\partial }{\partial r}\left(\frac{v_{\phi}}{r} \right)\right]^{2}.
\end{eqnarray}
The Compton cooling/heating rate 
is described as
\begin{eqnarray}
  \Gamma_{\rm Comp}&=&4\sigma_{\rm T}c\frac{k_{\rm B}\left(T_{\rm gas}-T_{\rm rad}\right)}{m_{\rm e}c^{2}}\left(\frac{\rho}{m_{\rm p}} \right)E_{0}.
\end{eqnarray}
Here, $m_{\rm e}$ is the electron mass and
$T_{\rm rad}\equiv (E_{0}/a)^{1/4}$ is the radiation temperature 
with $a=4\sigma_{\rm SB}/c$ being the radiation constant. 
Under the FLD approximation, $F_0$ and $\bm{{\rm P}}_{0}$ are calculated in terms of $E_0$. 

\begin{table*}[t]
  \tbl{Model parameters}{
  \begin{tabular}{lll}
      \hline
      parameter & symbol & value(s) \\
      \hline \hline
      black hole mass & $m_{\rm BH}\equiv M_{\rm BH}/M_{\odot}$  & $10,~10^4,~10^7$   \\
      mass injection rate & ${\dot m}_{\rm input} \equiv {\dot M}_{\rm input}/(L_{\rm Edd}/c^2)$ & $1000$, $2000^\ast$\\
      viscosity parameter & $\alpha$ & 0.1 \\
      simulation box: inner radius & $r_{\rm in}/r_{\rm S}$ & 2 \\
      simulation box: outer radius & $r_{\rm out}/r_{\rm S}$ &  6000 \\
      Keplerian radius & $r_{\rm K}/r_{\rm S}$ & 1000 \\ \hline
  \end{tabular}}
  \begin{tabnote}
    {$\ast$ We set ${\dot m}_{\rm input} = 1000$, except for model m1-high, in which we set ${\dot m}_{\rm input} = 2000$ (see table 2 for model description). }
  \end{tabnote}
  \label{parameter}
\end{table*}

\subsection{Initial condition and calculated model}

All the parameter values of calculated models are summarized in table \ref{parameter}.
We adopt a large Keplerian radius ($r_{\rm K} = 1000~r_{\rm S}$) and large computational box size $r_{\rm out}=6000~r_{\rm S}$.
We only solve the upper-half domain above the equatorial plane; i.e., $0 \leq \theta \leq\pi/2$.
The black hole mass and mass injection rate are set to be $m_{\rm BH}=  10,~ 10^{4}$ and $10^{7}$, and $\dot{m}_{\rm input}=\dot{M}_{\rm input}/(L_{\rm Edd}/c^2) =1000$, respectively.
We perform another simulation with $(m_{\rm BH},~ \dot{m}_{\rm input})=(10,~2000)$ to investigate a high $\dot{m}_{\rm BH}$ case with $m_{\rm BH}=10$ (named as model m1-high).

We initially put a hot optically thin atmosphere, which is assumed to be in uniform temperature distribution and hydrostatic equilibrium in the radial direction.
Then, the density profile is
\begin{eqnarray}
  \rho_{\rm atm}(r,\theta)&\equiv&\rho_{\rm out}\exp\left[\frac{\mu m_{\rm p}GM_{\rm BH}}{k_{\rm B}T_{\rm atm} r_{\rm out}}\left(\frac{r_{\rm out}}{r}-1\right)\right].
\end{eqnarray}
where $\rho_{\rm out}$ is the density at the outer boundary and $T_{\rm atm}$ is the temperature of hot optically thin atmosphere.
We employ $\rho_{\rm out}=10^{-17}{\rm g~cm^{-3}}$ and $T_{\rm atm}=10^{11}{\rm K}$, following \citet{ohsuga2005}.
We also assume that the initial radiation energy is $1/1000$ of the internal energy.

Matter is injected continuously at a constant rate of $\dot{m}_{\rm input}$
through the outer disk boundary at $r=r_{\rm out}$ and $0.49\pi\leq\theta\leq0.5\pi$.
The injected gas, with a temperature of $10^4 ~{\rm K}$, is assumed to possess a specific angular momentum corresponding to the Keplerian radius.
We thus expect that inflow material first falls towards the center
and forms a rotating gaseous ring around a radius of $r\sim r_{\rm K}$, 
from which the material slowly accretes inward via viscous diffusion process.

We assume that matter freely goes out but not comes in through the outer boundary ($r=r_{\rm out}$, $\theta$=$0 - 0.49\pi$) and the inner boundary ($r=r_{\rm in}$). 
For radiation, we apply $F_0^r=cE_0$ at the outer boundary and $F_0^r = - cE_0$ at the inner boundary, 
which means that the radiation energy goes freely outside the computational domain.

Grid points are uniformly distributed in logarithm in the radial direction;
$\triangle\log_{10} r = (\log_{10} r_{\rm out}-\log_{10} r_{\rm in})/N_{r}$, where $N_r$ is the number of grid points to be specified later,
while it is uniformly distributed in $\cos\theta$ in the polar direction;
$\triangle\cos \theta=1/N_{\theta}$, where the numbers of grid points are $N_{\theta}=240$.

To save computational times, we first set the inner boundary to be $r_{\rm in}=20~r_{\rm S}$ and $N_r=80$ (we call this a first-step simulation).
We start simulations with a nearly empty space around a black hole by setting a hot optically thin atmosphere with negligible mass.
After the gas stream reaches the innermost region and the gas inflow undergoes a transition to the super-Eddington state,
we start a second-step simulation by changing $N_r = 200$ and $r_{\rm in} = 2 ~r_{\rm S}$, using the interpolated first-step data.
In the following section we analyze and discuss the results of the second-step simulations.

\section{Results}
\subsection{Overall flow structure}
\label{sec-overall}


\begin{figure*}[]
  \begin{center}
    \includegraphics[width=110mm]{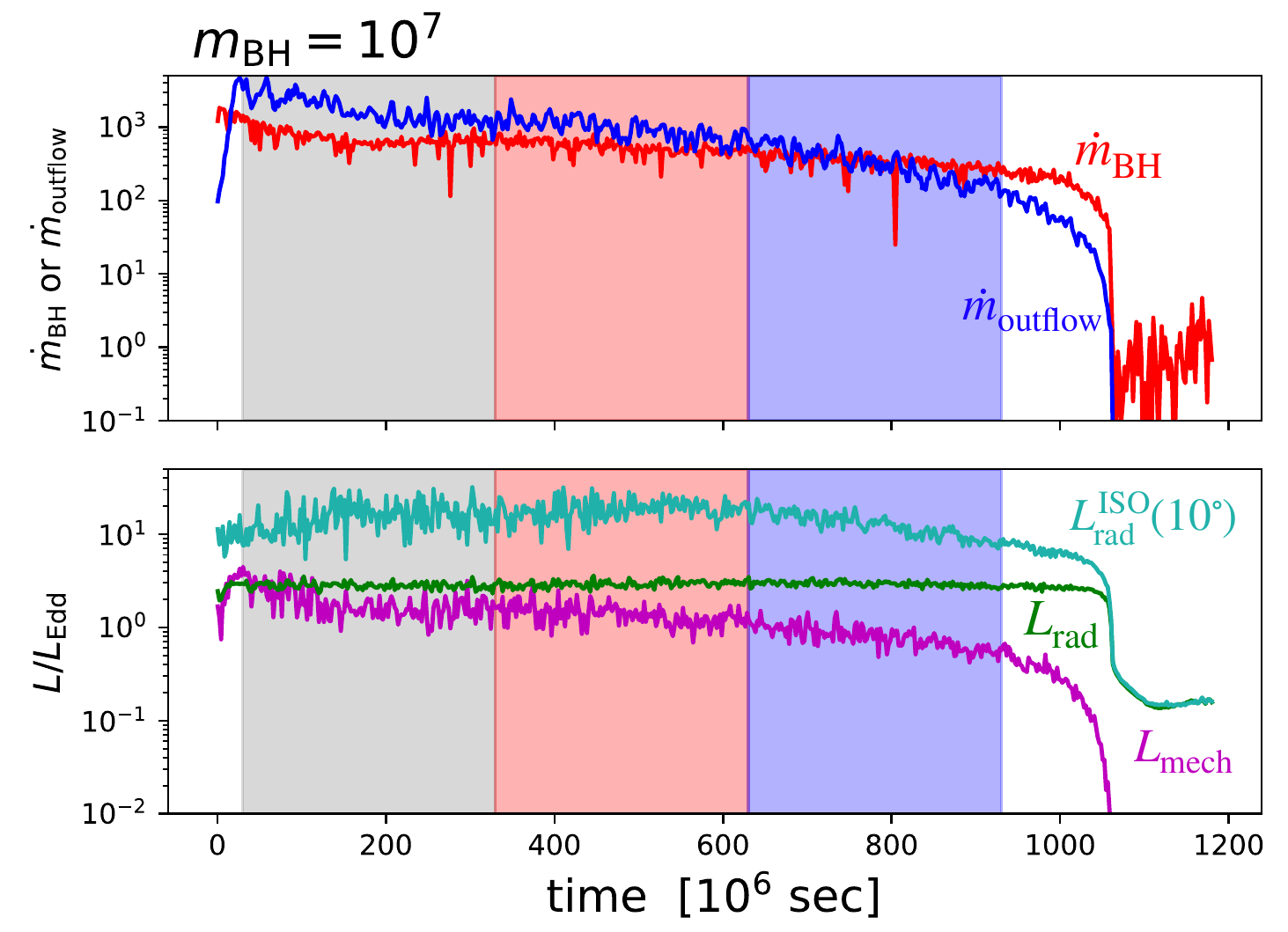}    
  \end{center}
  \caption{
    Time evolution of the normalized accretion rate (red), the normalized mass outflow rate (blue), 
    the normalized total mechanical luminosity (magenta), the normalized total radiation luminosity (green) and the normalized isotropic radiation luminosity (cyan). 
    The time $t$ shown in this figure represents that redefined in the second-step simulation.: 
    The time $t=0$ in the second-step simulation corresponds to $t=4420 \times 10^6$ sec in the first-step one.

  }
  \label{fig1}
\end{figure*}

\begin{table*}[t]
  \tbl{Radiation and outflow properties}{
  \begin{tabular}{llcccccc}
      \hline
      model& $m_{\rm BH}$  & $\dot{m}_{\rm BH}$ &  $\dot{m}_{\rm outflow}$ & $L_{\rm rad}/L_{\rm Edd}$& $L_{\rm mech}/L_{\rm Edd}$ & $L_{\rm rad}^{\rm  ISO}/L_{\rm Edd}$   &  $L_{\rm mech}^{\rm  ISO}/L_{\rm Edd}$ \\ 
      name &  &  &  &  &   &  ($\theta=10^\circ$) & ($\theta=10^\circ$) \\
      \hline \hline
      m1p1 (early phase) & $10$ & 280 & 88 & 2.6 & 0.4 & 7.5 & 5.7 \\
      m1p2 (mid phase) & $10$ & 300 & 110 & 2.6 & 0.4 & 7.5 & 5.7 \\
      m1p3 (late phase) & $10$ & 350 & 170 & 2.6 & 0.5 & 9.2 & 8.1 \\\hline
      m4p1 (early phase) & $10^4$ & 500 & 690 & 2.7 & 0.85 & 17 & 8.0 \\
      m4p2 (mid phase) & $10^4$ & 410 & 340 & 2.8 & 0.79 & 12 & 8.1 \\
      m4p3 (late phase) & $10^4$ & 350 & 200 & 2.7 & 0.53 & 9.2 & 6.5 \\ \hline
      m7p1 (early phase) & $10^7$ & 730 & 1900 & 2.7 & 1.5 & 14 & 12 \\
      m7p2 (mid phase) & $10^7$ & 550 & 1100 & 2.5 & 1.2 & 20 & 7.8 \\
      m7p3 (late phase) & $10^7$ & 370 & 390 & 2.8 & 0.7 & 12 & 7.3 \\ \hline \hline
      m1-high & $10$ & 480 & 550 & 2.7 & 0.86 & 18 & 8.1 \\ \hline
  \end{tabular}}
  \begin{tabnote}
   Here, 
   $\dot{m}_{\rm outflow}$ is the normalized outflow rate at $r_{\rm out}$, $L_{\rm rad}/L_{\rm Edd}$ is the normalized radiation luminosity, 
   $L_{\rm mech}/L_{\rm Edd}$ is the normalized mechanical luminosity, $L_{\rm rad}^{\rm  ISO}/L_{\rm Edd}$ is the normalized isotropic radiation luminosity, 
   and $L_{\rm mech}^{\rm  ISO}/L_{\rm Edd}$ is the normalized isotropic mechanical luminosity, respectively.
  \end{tabnote}
  \label{tab2}
\end{table*}

\begin{figure*}[]
  \begin{center}
    \includegraphics[width=160mm]{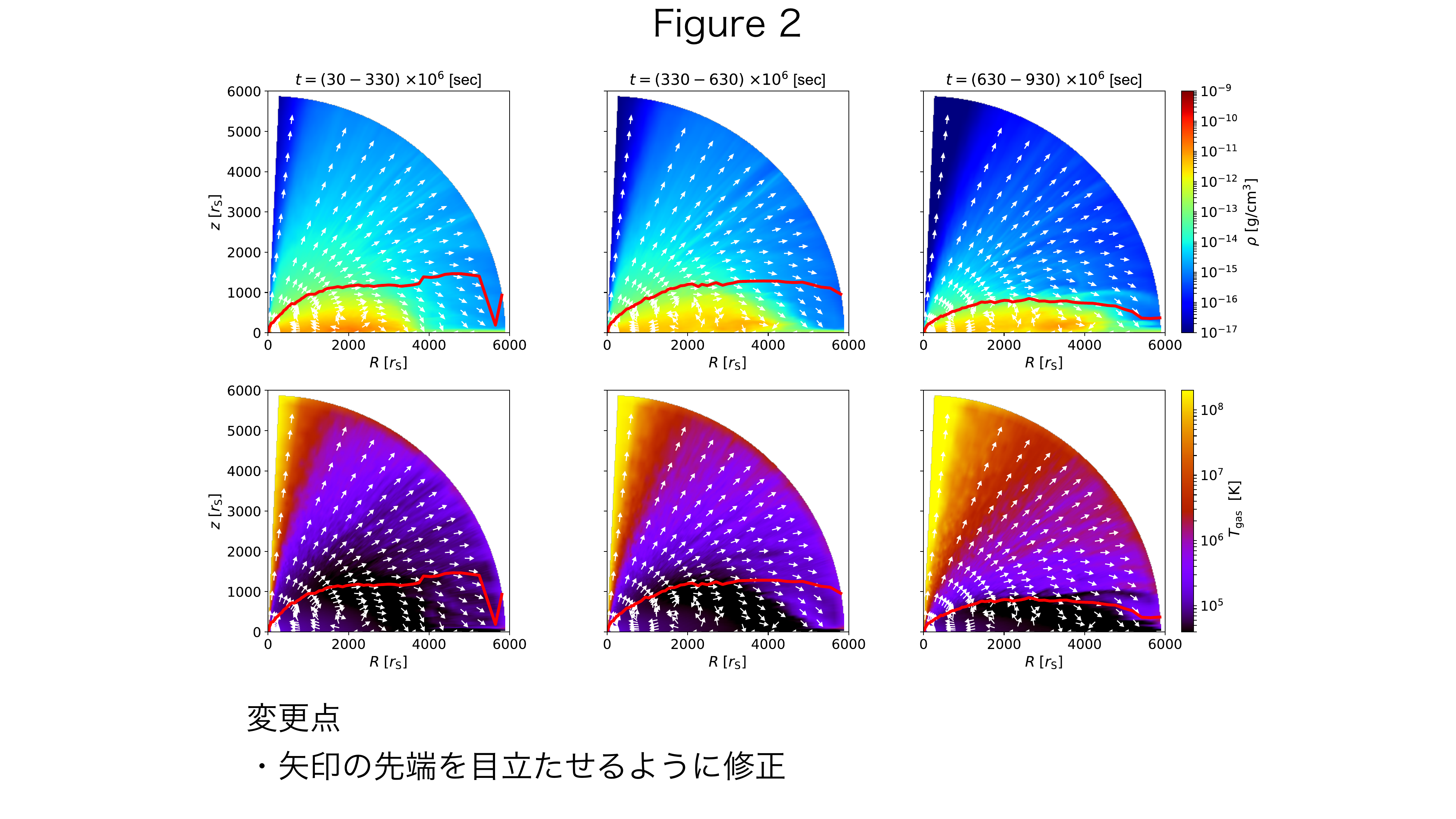}
  \end{center}
  \caption{
    Time-averaged density distributions (upper) and gas temperature distributions (lower)  
    around a super-massive black hole in m7p1 (left), m7p2 (center) and m7p3 (right), respectively.
    The time-averaged interval is $ 300 \times 10^6 ~{\rm sec}$.
    Overlaid are the directions of the gas velocity vectors and the disk surface (red solid line) which is defined as the loci where the radiation force balances the gravitational force.
  }
  \label{fig2}
\end{figure*}

As mentioned in section 2, we start a first-step simulation by setting $r_{\rm in } = 20~ r_{\rm S}$.
The first-step simulation is displayed in the Appendix.
The pattern of luminosity variation is almost the same for both 1st and 2nd step, as the luminosity and the disk structure fluctuate due to the disk instability outside the inner boundary of the 1st step simulations ($r_{\rm in}=20~r_{\rm S}$).
However, $L_{\rm mech}/L_{\rm Edd}$ and $L_{\rm rad}/L_{\rm Edd}$ increase from 0.2 to 1.5 and from 2.4 to 2.6, respectively.
The reason $L_{\rm rad}/L_{\rm Edd}$ does not increase significantly is due to the photon trapping effect, where photons inside $r=20~r_{\rm S}$ are absorbed by the black hole.
We notice in most cases that the flow undergoes high-low transitions 
in luminosities (figure \ref{figa1} in the Appendix) as a result of the thermal instability 
of a sort found in super-Eddington accretion flow (see, e.g., Honma et al. \citeyear{honma1991}, see also Kato et al. \citeyear{kato2008}, Chap. 10, for a concise review and references therein).
We start a second-step simulation by using the data just after upward transition is completed.

More precisely, we use the data just after the onset of the first upward transition in the cases of $m_{\rm BH} = 10$ and $10^7$, while in the case of $m_{\rm BH} =10^4$ we use the data just after the onset of the fourth upward transition (see gray line in figure \ref{figa1} in the Appendix).
Since the overall structure of the accretion flows and outflows are qualitatively the same in our parameter space of the black hole mass, 
we only demonstrate the overall structure of models with $m_{\rm BH} = 10^7$ in this subsection. 

We first show in figure \ref{fig1} time evolution of accretion rate and luminosities for the second-step simulation of the case of $m_{\rm BH}=10^7$.
Hereafter, we reset the time t=0 at the beginning of the second-step simulation, where this time corresponds to $t=4420 \times 10^6 $ sec.
The red, blue, magenta, green and cyan lines represent the normalized accretion rate, outflow rate, total mechanical luminosity, total radiation luminosity and isotropic radiation luminosity, respectively.
Here, we calculate the normalized  accretion rate, outflow rate, 
\begin{eqnarray}
  \dot{m}_{\rm BH}~ (r=5~r_{\rm S})&\equiv&4\pi \frac{c^2}{L_{\rm Edd}} \int_{\theta_{\rm surf}}^{\pi/2} d\theta\sin\theta\nonumber\\
  &&~~~~~~\times r^{2}\rho(r,\theta){\rm min}\left\{v_{r}(r,\theta),0\right\},\\
  \dot{m}_{\rm outflow}~ (r=r_{\rm out})&\equiv&4\pi\frac{c^2}{L_{\rm Edd}} \int_{0}^{\theta_{\rm surf}} d\theta\sin\theta\nonumber\\
  &&~~~~~~\times r^{2}\rho(r,\theta){\rm max}\left\{v_{r}(r,\theta),0\right\}.   \label{eq-mdotoutflow}
\end{eqnarray}
where, $\theta_{\rm surf}=\theta_{\rm surf}(r)$ is the $\theta$-coordinate of the disk surface (see the red line in figure \ref{fig2}).
We also calculate the radiation and mechanical luminosities measured at $r=r_{\rm out}$ by
\begin{eqnarray}
&L_{\rm rad}& =4\pi \int_{0}^{\theta_{\rm surf}} \ 
r^2 {\rm max}\{F^{r}_{\rm lab},\ 0\} \sin{\theta} d\theta, \\ 
&L_{\rm mech}&=4\pi\int_{0}^{\theta_{\rm surf}}
 r^2 ~{\rm max}\{\frac{1}{2}\rho v^2 v_r,\ 0 \}\sin{\theta} d\theta,\\
&L_{\rm rad}^{\rm  ISO} (\theta)& =4\pi r^2 ~{\rm max} \{F^{r}_{\rm lab},\ 0\},\\
&L_{\rm mech}^{\rm  ISO}(\theta)&=4\pi r^2\  {\rm max}\{\frac{1}{2}\rho v^2 v_r,\ 0\}.
\end{eqnarray}
Here, $v^2 = v_{\rm r}^2 + v_{\rm \theta}^2 + v_{\rm \phi}^2$
and $F_{\rm lab}^{r}$ is the radial component of radiation flux in the laboratory frame\ (see K21).
In K21 and Paper I, we used the notations of $L_{\rm X}$ and $L_{\rm X}^{\rm ISO}$ for representing the total and isotropic radiation luminosities, but here we use the notations of $L_{\rm rad}$ and $L_{\rm rad}^{\rm  ISO}$ in this paper because the radiation is converted to the UV band in the cases of super-massive black holes.
As is clearly seen in figure \ref{fig1}, 
the accretion rate and luminosity exhibit large-amplitude variations and
the transition from the super-Eddington to the sub-Eddington state.
Therefore, we analyzed the super-Eddington state in three phases (gray, red, blue), named as m7p1, m7p2, and m7p3, respectively, from the early to later phases.
We summarized the averaged values of normalized accretion rate and luminosities for each phase in table \ref{tab2}.

We next show in figure \ref{fig2} density (upper) and gas temperature (lower) distributions overlaid with the velocity fields for m7p1 (left), m7p2 (center) and m7p3 (right), respectively.
The disk surface is defined as the loci where radiation force balances the gravity in radial direction, $\chi F_{0,r} /c = \rho GM_{\rm BH} /(r - r_{\rm S} )^2 $.
Comparing the left (early) and right (late) panels, we notice that the disk (red line) becomes geometrically thin as the normalized accretion rate decreases.
Furthermore, a large amount of low-temperature outflow ($\sim 10^5~ {\rm K}$) is launched in the region far away from the black hole ($ R \gtrsim 1000~ r_{\rm S}$) in the early/mid phase (m7p1, m7p2), while high-temperature outflow ($ \gtrsim 10^7~ {\rm K}$) is launched with a large opening angle ($\theta \sim 30^\circ$) from the vicinity of the black hole in the late phase (m7p3).

\subsection{Radiation and mechanical luminosities}
\label{sec-mdot-L}
\begin{figure*}[t]
  \begin{center}
    \includegraphics[width=100mm]{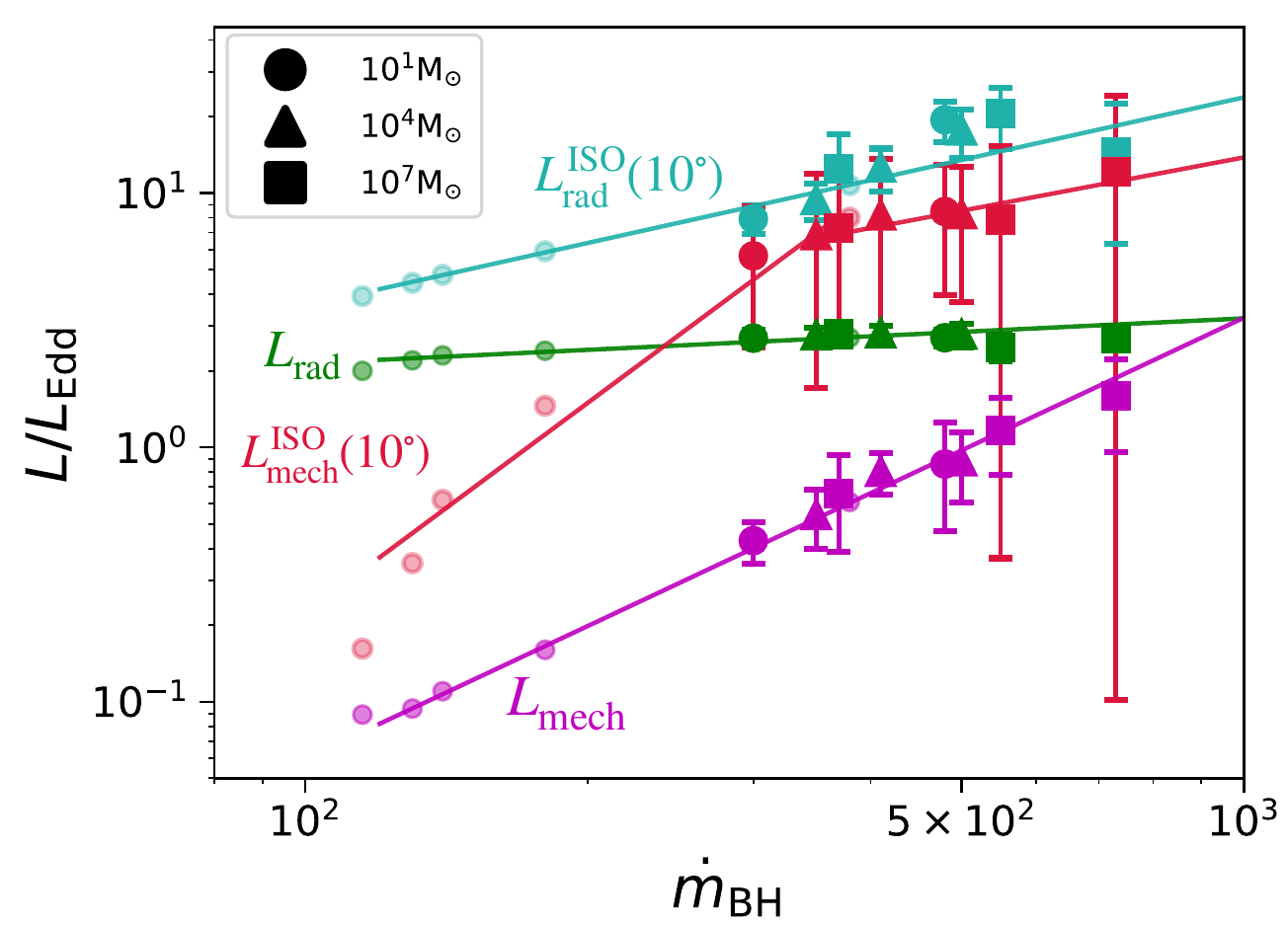}
  \end{center}
  \caption{
   Radiation and mechanical luminosities as a function of $\dot{m}_{\rm BH}$.
   The circle, triangle and square represent $m_{\rm BH} = 10^1,~10^4$ and $10^7$, respectively.
   The magenta, green, red and cyan represent total mechanical, total radiation, isotropic mechanical and isotropic radiation luminosity, respectively.
  }
  \label{fig3}
\end{figure*}
Figure \ref{fig3} shows the radiation and mechanical luminosities normalized by $L_{\rm Edd}$ as functions of $\dot{m}_{\rm BH}$.
The thin plot is the result of Paper I.
For calculating isotropic luminosities we assume the viewing angle of $\theta = 10^\circ$.

We first notice that the $\dot{m}_{\rm BH}$ dependence of normalized luminosity found by Paper I for stellar mass BHs still holds for total luminosity (green and magenta) and isotropic radiation luminosity (cyan).
In contrast, the normalized isotropic mechanical luminosity does not hold for high normalized accretion rates ($\dot{m}_{\rm BH} \gtrsim 400$) and show broken power-law.
At low $\dot{m}_{\rm BH}$ ($\dot{m}_{\rm BH} < 400$), the normalized isotropic mechanical luminosity rapidly grows in proportion to $\dot{m}_{\rm BH}^{2.7}$ as $\dot{m}_{\rm BH}$ increases, 
while at high normalized accretion rates it increases more slowly ($\propto \dot{m}_{\rm BH}^{0.7}$).
Comparing the plots of $10^4 ~M_{\odot}$ and $10^7 ~M_{\odot}$ for $\dot{m}_{\rm BH}\sim 370$ (m4p3 and m7p3), 
we understand that the normalized luminosities are independent of black hole mass.
Furthermore, figure \ref{fig3} shows that the fitting formula holds for all mass ranges ($10^1 - 10^7 M_\odot$) in the case of high normalized accretion rates ($\dot{m}_{\rm BH} \gtrsim 400$).
We conclude that the relationship between the normalized luminosities and the normalized accretion rate does not depend on the black hole mass.

\subsection{Why broken power-law?}
\begin{figure}[t]
  \begin{center}
    \includegraphics[width=60mm]{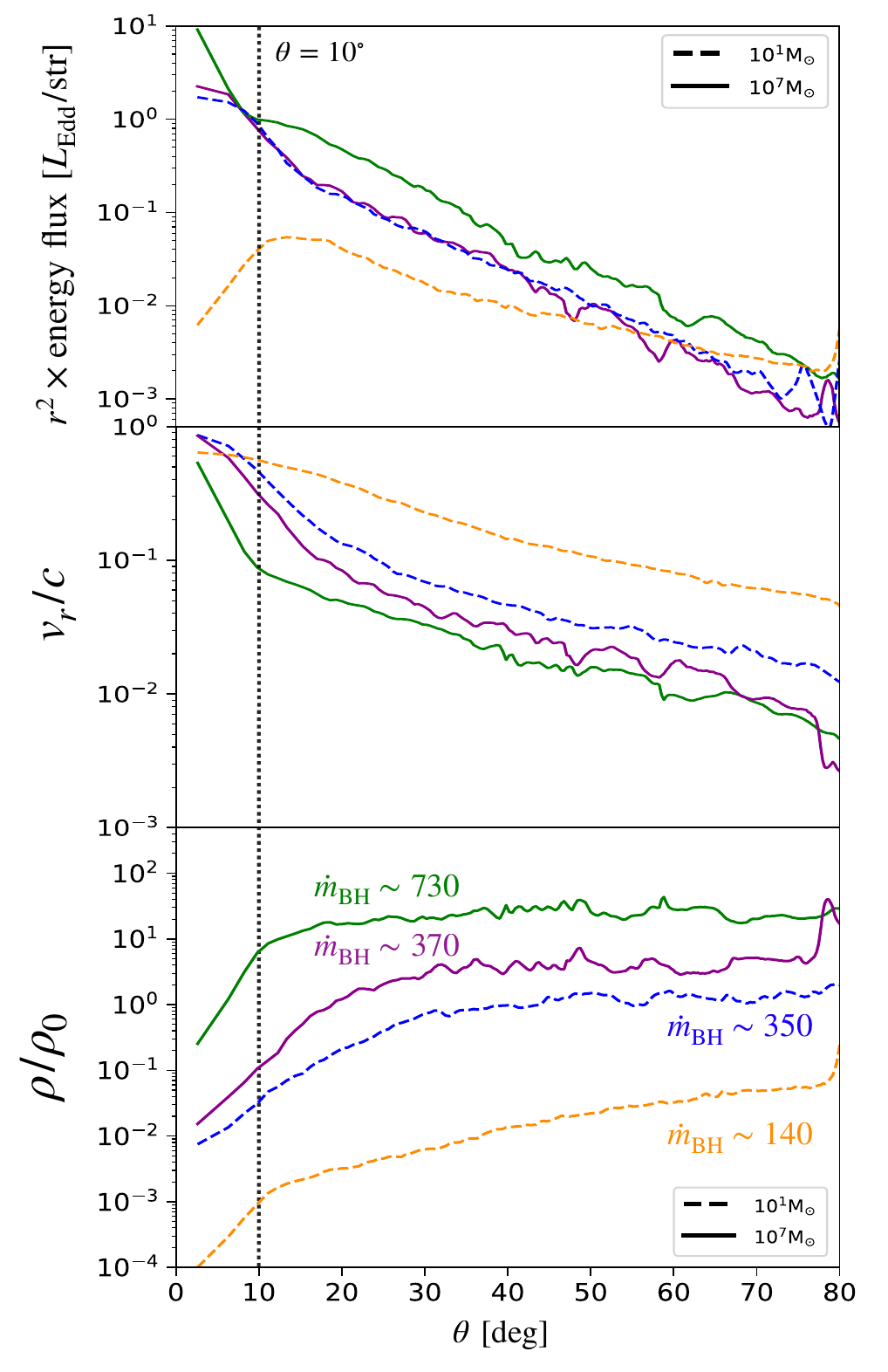}
  \end{center}
  \caption{
    [Top] The angular distribution of mechanical energy flux, with solid and dashed lines representing $m_{\rm BH} = 10^1$ and $10^7$.
          The green, purple, blue, and orange lines represent m7p1 ($\dot{m}_{\rm BH} \sim 730$), m7p3 ($\dot{m}_{\rm BH} \sim 370$), m1p3 ($\dot{m}_{\rm BH}\sim 350$) and Model-140 in Paper I ($\dot{m}_{\rm BH}\sim 140$), respectively. 
    [Middle] Same as the top panel but for $v_r/c$. 
    [Bottom] Same as the top panel but for $\rho/\rho_0$. 
  }
  \label{fig4}
\end{figure}

\begin{figure*}[t]
  \begin{center}
    \includegraphics[width=160mm]{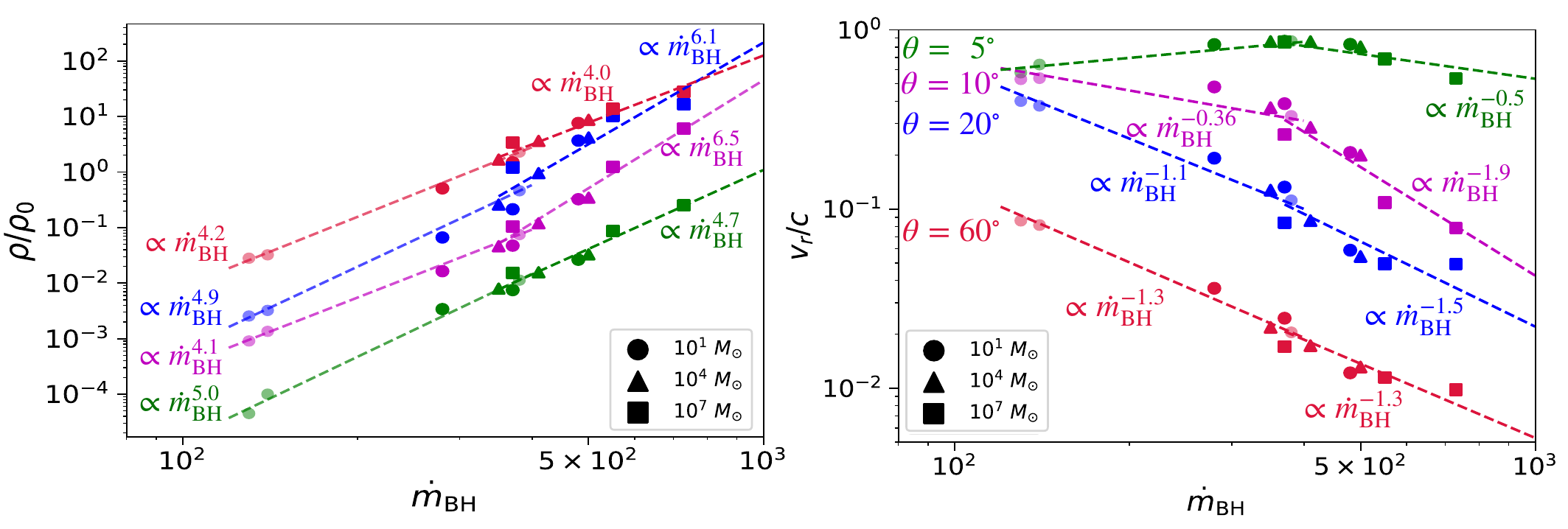}
  \end{center}
  \caption{
    The $\dot{m}_{\rm BH}$-dependence of $\rho$ (left) and $v_r$ (right) at $r_{\rm out}$ for $\theta=5^\circ$ (green), $10^\circ$ (magenta), $20^\circ$ (blue) and $60^\circ$ (red), respectively.
    The thin plots are the results of Paper I.
  }
  \label{fig6}
\end{figure*}



Why does the normalized isotropic mechanical luminosity exhibit a broken power-law relationship with respect to the normalized accretion rate? 
To explore the reason, we plot the angular dependence of the normalized mechanical energy flux, gas velocity, and 
density for each $\dot{m}_{\rm BH}$ in figure \ref{fig4}.
The normalized density values $\rho_0$ for $m_{\rm BH}=10, 10^4,$ and $10^7$ are $10^{-10}$, $10^{-13}$, and $10^{-16}$ [$\rm g/cm^3$], respectively.
Let us pay attention to the shape of each line near the polar axis ($\theta = 10^\circ$). 
We notice that the normalized mechanical energy flux rapidly increases with increasing normalized accretion rate.
Comparing the normalized density and velocity distributions at $\theta=10^\circ$ for m7p1 (green) and m7p3 (purple), we see that the normalized density is 66 times larger when $\dot{m}_{\rm BH}$ is doubled, while the velocity decreases to 1/3.
When we compare the cases of $\dot{m}_{\rm BH}=350$ and $140$, we notice that the normalized density has increased by a factor of 35, while the velocity has changed by $\sim 20~\%$ only, around $0.5c$ in both cases.

Figure \ref{fig4} clearly shows the existence of two components of outflow; i.e., less massive and faster ($> 0.1 c$) component inside and more massive and slower ($< 0.1 c$) component outside.
From the angular distribution of velocity, we find that the faster component of outflow is widely ejected over $0^\circ$ to $60^\circ$ at low $\dot{m}_{\rm BH}$, whereas the faster component is only ejected in the polar direction at high $\dot{m}_{\rm BH}$.
Likewise, the angular profile of density also exhibits two components: it is flat at large $\theta$ (towards the equatorial plane), while it shows a sharp decrease as $\theta$ vanishes (towards the polar direction).
The critical angle which separates the two components tends to decrease, as $\dot{m}_{\rm BH}$ increases and is $\sim 20^\circ$ (or $\sim10^\circ$) in models with $\dot{m}_{\rm BH}\sim 370$ (purple) (in model with $\dot{m}_{\rm BH}\sim 730$). 
We understand that the presence of the two component outflow is responsible for the broken power-law type relationship between ${\dot m}_{\rm BH}$ and the isotropic mechanical luminosity.

Let us illustrate the same but in a different way.
Figure \ref{fig6} shows the $\dot m_{\rm BH}$-dependence of normalized density (left) and velocity (right) 
for given polar angles; $\theta =$ 5, 10, 20, and 60 ($^\circ$), respectively.
We find that the $\dot{m}_{\rm BH}$ dependence of the density (shown in the left panel of figure \ref{fig6}) 
and the velocity (right panel of figure \ref{fig6}) shows a broken power-law distribution with a break at around $\dot{m}_{\rm BH}\sim 400$.
If we see the line of $\theta =10^\circ$, 
we notice that with the increase of $\dot{m}_{\rm BH}$, the normalized density increases in proportion to $\dot{m}_{\rm BH}^{6.5}$ ($\dot{m}_{\rm BH}^{4.1}$) above (below) $\dot{m}_{\rm BH}\sim 400$, 
whereas the normalized velocity decreases only slowly ($\propto \dot{m}_{\rm BH}^{-0.36}$) at low $\dot{m}_{\rm BH}$, 
but steeply decreases ($\propto \dot{m}_{\rm BH}^{-1.9}$) above $\dot{m}_{\rm BH}\sim 400$.
That is, around $\dot m_{\rm BH}\sim 400$, the power of $\dot m_{\rm BH}$-dependence of normalized density increases only by a factor of 1.5, while the power of $\dot m_{\rm BH}$-dependence of velocity increases by a factor of 5 from -0.36 to -1.9.
To conclude, the reason for the broken power-law relationship arises due to a change of $\dot{m}_{\rm BH}$ dependence of the normalized radial velocity across $\dot{m}_{\rm BH} \sim 400$. 

We may thus conclude that slower and denser outflows are ejected in the poleward direction, when the accretion rate is high. This is because the disk vertically inflates and thus collimates the outflow.
As a result, density rapidly increases, whereas radial velocity rapidly decreases, as ${\dot m}_{\rm BH}$ increases.

\section{Discussion}
\subsection{Why are the normalized luminosities insensitive to black hole mass?}
 
Figure \ref{fig3} clearly demonstrates that both of the normalized radiation and mechanical luminosities increase with an increase of the normalized accretion rate, 
but that $L_{\rm rad}/L_{\rm Edd}$ and $L_{\rm mech}/L_{\rm Edd}$ are rather insensitive to the black hole mass.
This is not surprising for the following reasons.
First, the normalized radiation luminosity can be written simply as 
$L_{\rm rad}/L_{\rm Edd} = \eta_{\rm rad} \dot{m}_{\rm BH}$
with $\eta_{\rm rad}$ being the radiation efficiency (which is a function of the normalized accretion rate, $\dot m_{\rm BH}$). 
Therefore, $L_{\rm rad}/L_{\rm Edd}$ should be insensitive to the black hole mass,
as long as $\dot{M}_{\rm BH}$ is normalized by $L_{\rm Edd}/c^2$.

Then, how about the mechanical luminosity?
The super-Eddington accretors are accelerated by radiation force, which is expressed in terms of opacity and radiation flux.
As discussed above, the normalized radiation flux 
[with a normalization factor of $ L_{\rm Edd}/(4\pi r_{\rm S}^2)\propto M_{\rm BH}^{-1}$]
does not depend on the black hole mass. 
In addition, opacity $\kappa ~ (= \chi/\rho)$ does not depend on the black hole mass, either,
since electron scattering dominates over absorption in the whole region. 
Thus, the acceleration does not depend on the black hole mass.

These are naive explanations why the normalized total luminosities (both of radiation and mechanical ones)
does not show $m_{\rm BH}$ dependences, however, how to understand the behavior of their isotropic values?
In other words, why are the angular profiles of radiation and mechanical flux the same for the same ${\dot m_{\rm BH}}$.
We numerically confirm that for the same accretion rate, the scale-height ($H$) of the accretion flow normalized by $r_{\rm S}$ as a function of $r/r_{\rm S}$ remains nearly constant, even if we vary $m_{\rm BH}$.
Since the direction of normalized radiation flux should depend on the geometrical shape of the photosphere,
this would indicate that the radiation flux profile is ${\dot m}_{\rm BH}$ independent,
so is the mechanical flux profile.

However, the effect of opacity is also important when considering super-massive black holes.
As discussed in \citet{jiang2020}, 
the iron opacity bump causes convective instability when it exceeds the opacity of the electron scattering. 
As a result, the disk structure and luminosity fluctuate.
In particular, luminosity fluctuation is three times larger when convection is taken into account, comparing the case of normal MRI turbulence without convection and iron opacity bump.
The effect of opacity depends on density and temperature, suggesting that it depends on black hole mass. 
Therefore, the dependence of normalized luminosity on the normalized accretion rate may vary with black hole mass.
It is also a future issue to incorporate the effects of opacity in this way.

\subsection{Is the simulation box large enough?}

\begin{figure*}[t]
  \begin{center}
    \includegraphics[width=80mm]{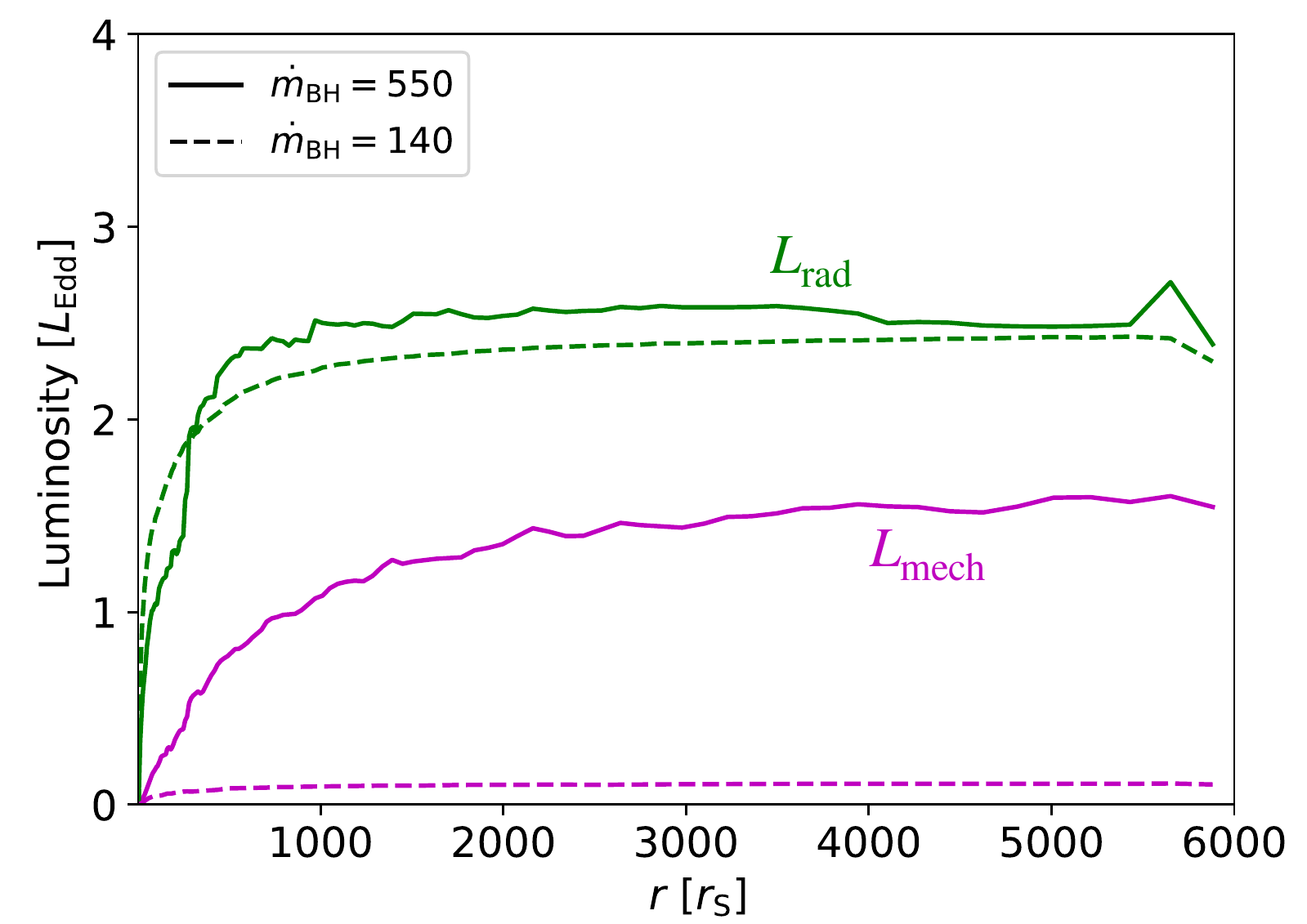}
  \end{center}
  \caption{
    The radial distribution of total radiation (green) and mechanical (magenta) luminosities of $\dot{m}_{\rm BH} \sim 550$ (solid) and $\dot{m}_{\rm BH} \sim 140$ (dashed), respectively.
  }
  \label{r_lum}
\end{figure*}

To quantitatively evaluate AGN feedback, it is necessary to understand the radial dependence of energy fluxes.
Figure \ref{r_lum} shows the radial distribution of total radiation (green) and mechanical (magenta) luminosities of $\dot{m}_{\rm BH}\sim 550$ (solid) and $\dot{m}_{\rm BH}\sim 140$ (dashed), respectively.
We find that the radial dependence of normalized radiation luminosity is flat at a few thousand $r_{\rm S}$, independent of normalized accretion rate.
In contrast, 
the radial dependence of the normalized total mechanical luminosity varies with the normalized accretion rate, showing a flat profile at a few thousand $r_{\rm S}$ at low normalized accretion rates but appearing to increase slowly at high normalized accretion rates, even at a few thousand $r_{\rm S}$.
\citet{botella2022} performed axisymmetric two-dimensional RHD simulation using a nested simulation-box method to investigate large-scale ($\sim 10^6~r_{\rm S}$) structure of outflows.
\citet{botella2022} suggests that the normalized mechanical energy flux in the $45^\circ~(80^\circ)$ direction increases to $10^4~(10^5)~r_{\rm S}$, 
so that physical quantities calculated at a few $1000~r_{\rm S}$ can not simply be extrapolated.
Simulation over a larger region is to be attempted in future work.

\subsection{Application to observations}

\begin{table*}[t]
  \tbl{luminosities of NLS1s}{
  \begin{tabular}{cllll}
      \hline
      object  &$m_{\rm BH}$ & $L_{\rm rad}^{\rm  ISO}(\theta)/L_{\rm Edd}$  & $L_{\rm mech}^{\ast}(\theta)/L_{\rm Edd}$\footnotemark[$\dag$]
      &  $L_{\rm mech}^{\ast}/L_{\rm rad}^{\rm  ISO}(\theta)$\footnotemark[$\dag$]\\
      \hline 
      1H 0707$-$495 & $ 4\times 10^6 $  & $\sim 1.2 $ &  $\sim 0.11$ & $\sim 0.09$ \\
      1H 0323$+$103 & $2\times 10^{7}$ & $\sim 1$ & $\sim 0.05$ &  $\sim 0.05$   \\\hline
      simulation & $10$ & $2.4-4.8$ & $\sim 0.11 ~ (0.02-0.56)$ &  $0.02-0.05$ $(0.02-0.12)$  \\
      $\left[ \dot{m}_{\rm BH} \sim 140 \right]$ & & ($\theta=0^\circ-\theta_{\rm surf}$)\footnotemark[$\ddag$] & ($\theta=0^\circ-\theta_{\rm surf}$)\footnotemark[$\ddag$] 
      & ($\theta=0^\circ-\theta_{\rm surf}$)\footnotemark[$\ddag$]  \\ \hline
  \end{tabular}}
  \begin{tabnote}
    \footnotemark[$\dag$] $L_{\rm mech}^{\ast} = L_{\rm mech}$ (1H 0707-495) or $L_{\rm mech}^{\rm ISO}$ (1H 0323+103) \\
    \footnotemark[$\ddag$] Here, ${\theta}_{\rm surf}$ represents the angle of the disk surface at $r_{\rm out}$, 
    which equals $80^\circ$ in Model-140 in Paper I ($\dot{m}_{\rm BH}\sim 140$).
  \end{tabnote}
  \label{obs_luminosity}
\end{table*}

\begin{figure*}[t]
  \begin{center}
    \includegraphics[width=170mm]{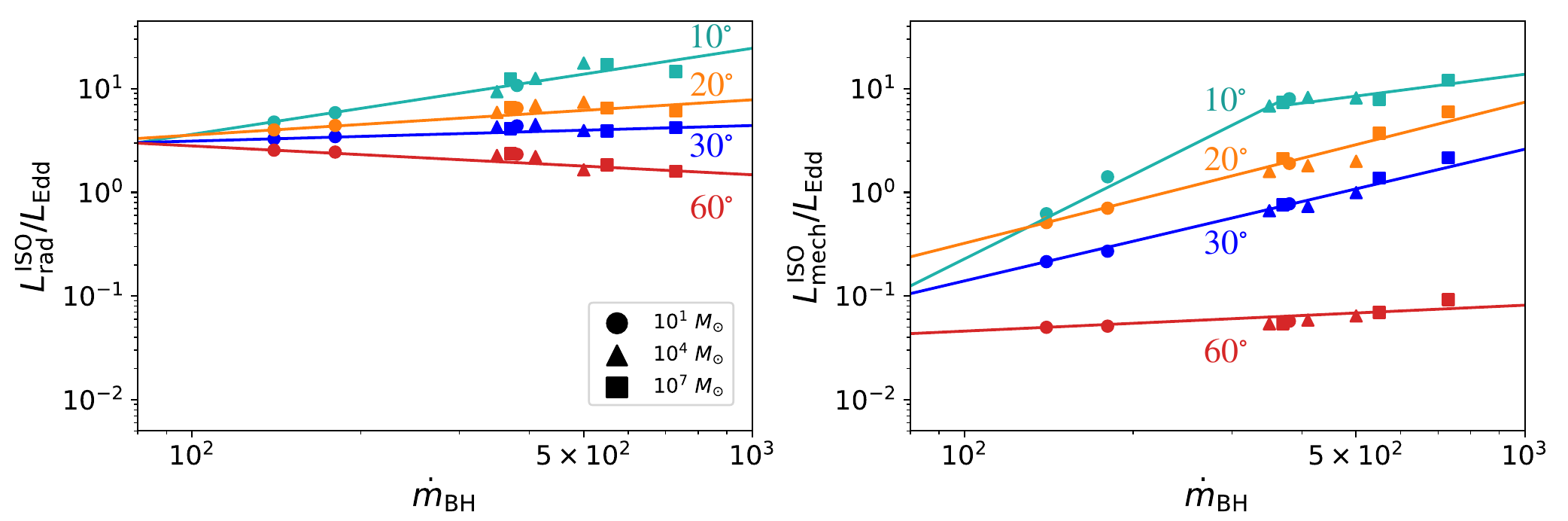}
  \end{center}
  \caption{
    Isotropic radiation (left) and mechanical (right) luminosities as functions of $\dot{m}_{\rm BH}$ for $10^\circ$ (cyan), $20^\circ$ (orange), $30^\circ$ (blue) and $60^\circ$ (red), respectively.
  }
  \label{fig7}
\end{figure*}


\subsubsection{Observations of NLS1}
Some NLS1s are known to be accompanied with high velocity outflow component with extents of 10 -- 100 parsecs \citep[e.g.,][]{reeves2003, tombesi2010}. 
In the cases of near or super-Eddington accretors, such outflows are likely to be radiation driven wind originating from super-Eddington accretion flow. 
The isotropic radiation luminosities and the mechanical luminosities of the outflow are listed in table \ref{obs_luminosity} for two NLS1s.
The radiation luminosities can reach $\sim L_{\rm Edd}$  at high ones \citep{done2016, jin2017, landt2017}. 
Measured mechanical luminosities are, by contrast, much less; only around $\sim 0.05 - 0.1~L_{\rm Edd}$.
The luminosity ratios are thus $(L_{\rm mech}/L_{\rm rad}^{\rm ISO},~ L_{\rm mech}^{\rm ISO}/L_{\rm rad}^{\rm ISO})\sim$ 0.05 -- 0.1 (see table 3), although there exist large uncertainties.
In this section, we discuss how the simulation results explain such observations.

\subsubsection{Predictions based on the simulation results}
Before making direct comparison, let us summarize what we know from our simulation results.
Figure \ref{fig7} show the normalized isotropic radiation (left) and mechanical (right) luminosities as functions of $\dot{m}_{\rm BH}$ for viewing angles of $10^\circ$ (cyan), $20^\circ$ (orange), $30^\circ$ (blue), and $60^\circ$ (red), respectively.
We notice that the dependence of normalized luminosity on the normalized accretion rate can be expressed by a single power-law, except for $L_{\rm mech}^{\rm ISO}(10^\circ)/L_{\rm Edd}$. 
Power-law indices are indicated in this figure \ref{fig7}.
Furthermore, we find that $\dot{m}_{\rm BH}$-dependence is somewhat flatter for the normalized radiation than for the normalized mechanical luminosity in the range of $i= 10^\circ - 30^\circ$.

In our simulations, we find that $L_{\rm mech}/L_{\rm Edd} \sim 0.1$ and $L_{\rm rad}^{\rm ISO}/L_{\rm Edd} \sim 2.4-4.8$ at low normalized accretion rates ($\dot m_{\rm BH}\sim 140$). 
We find that $L_{\rm mech}/L_{\rm rad}^{\rm ISO}$ is $\sim 0.02-0.05$ (see table 2 of Paper I). 

We also fitted the normalized luminosity ($L/L_{\rm Edd}$) for all simulation results
as functions of the normalized accretion rate, 
and obtained the following
fitting formulae for the normalized mechanical luminosity and isotropic radiation luminosities;
\begin{eqnarray}
  L_{\rm mech} / L_{\rm Edd} &=& 0.06 \left( \frac{\dot{m}_{\rm BH}}{100} \right)^{1.7}, \label{Lmech}\\
  L_{\rm rad}^{\rm ISO} (10^{\circ}) / L_{\rm Edd} &=& 3.6  \left( \frac{\dot{m}_{\rm BH}}{100} \right)^{0.83}, \\
  L_{\rm rad}^{\rm ISO} (20^{\circ}) / L_{\rm Edd} &=& 3.6  \left( \frac{\dot{m}_{\rm BH}}{100} \right)^{0.34}, \\
  L_{\rm rad}^{\rm ISO} (30^{\circ}) / L_{\rm Edd}  &=& 3.1  \left( \frac{\dot{m}_{\rm BH}}{100} \right)^{0.15}, \\
  L_{\rm rad}^{\rm ISO} (60^{\circ}) / L_{\rm Edd}  &=& 2.9  \left( \frac{\dot{m}_{\rm BH}}{100} \right)^{-0.37}.
\end{eqnarray}
Similarly, the fitting formulae for the normalized isotropic mechanical luminosities are:
\begin{eqnarray}
\label{Lmech-ISO}
  L_{\rm mech}^{\rm ISO} (10^{\circ}) / L_{\rm Edd} &=& 0.23  \left( \frac{\dot{m}_{\rm BH}}{100} \right)^{2.7}, \\ 
  L_{\rm mech}^{\rm ISO} (20^{\circ}) / L_{\rm Edd} &=& 0.32  \left( \frac{\dot{m}_{\rm BH}}{100} \right)^{1.4}, \\
  L_{\rm mech}^{\rm ISO} (30^{\circ}) / L_{\rm Edd} &=& 0.14  \left( \frac{\dot{m}_{\rm BH}}{100} \right)^{1.3}, \\
  L_{\rm mech}^{\rm ISO} (60^{\circ}) / L_{\rm Edd} &=& 0.04  \left( \frac{\dot{m}_{\rm BH}}{100} \right)^{-0.3}. 
\end{eqnarray}

In the following subsections we will use these formulae to compare with the observations of two NLS1.

\subsubsection{Case Study 1: 1H 0707-495}
Let us first compare the simulation with the observation of 1H 0707-495 (see table \ref{obs_luminosity}).
\citet{done2016} estimate the bolometric disk luminosity from the optical spectra (5100 {\AA}). 
\citet{xu2021} calculate the mechanical luminosity from the velocity and ionization parameters of outflows which arrive in the observer's direction,
as estimated from SED modeling of X-ray spectroscopy.
Therefore, we assume that isotropic mechanical luminosity is more appropriate to compare with the observations except that the solid angle of outflow. 
From the observation it is estimated to be $\Omega \sim 0.725$, whereas we assumed $\Omega = 4\pi$ in calculating $L_{\rm mech}^{\rm ISO}$.
We thus corrected the simulation value [see equation (\ref{Lmech-ISO})] by multiplying a factor of $4\pi/\Omega$.

Before comparison, we need to make a remark. 
Although our simulation results show that $L_{\rm rad}^{\rm ISO}$ exceeds the Eddington luminosity by a factor of $\sim 3$ in the super-Eddington regime, while the observation shows $L_{\rm rad}^{\rm ISO}/L_{\rm Edd}\sim 1$.
Possible reasons for such a discrepancy will be discussed in section \ref{sec4.3.5}.


We thus start with fitting to the isotropic mechanical luminosity.
To reproduce the observed (isotropic) mechanical luminosity of $L_{\rm mech}^{\rm ISO}/L_{\rm Edd} = 0.11$, 
we find ${\dot m}_{\rm BH} $ = (86, 57, 108, 29000) for 
$\theta$ = (10, 20, 30, 60), respectively, leading to the isotropic radiation luminosity of
$L_{\rm mech}/L_{\rm Edd} = (3.3,~3.0, ~3.2, ~0.4)$.

Then, the estimated normalized accretion rate is $45-86~(10^\circ-30^\circ)$ and the normalized isotropic radiation luminosity is  $0.84 - 3.1$. 
The isotropic luminosity ratio, is approximately, $L_{\rm mech}^{\rm ISO}/L_{\rm rad}^{\rm ISO} = 0.04 - 0.13$.
Further, the outflow velocity of 1H 0707-495 is suggested to be $0.15 c$ \citep{xu2021}.
We note that the case of $\theta = 60^\circ$ may not be appropriate, 
since the radiation luminosity calculated from the fitting formula is much higher than the observed accretion rate \citep[$ \dot{m}_{\rm BH} \sim 140 - 260$,][]{done2016}.
We hence suggest that the viewing angle should be moderate, $30^\circ - 60^\circ$, to explain the observed slow outflow, 
since figure \ref{fig4} shows that fast outflow erupts in the polar direction at low normalized accretion rates ($\dot m_{\rm BH}\leq 140$).
This result is consistent with \citet{done2016}, which also argued for high viewing angle ($30^\circ - 60^\circ$).

\subsubsection{Case Study 2: 1H 0323+103}
Let us next compare with the observation of 1H 0323+103 \citep[e.g.,][]{doi2012, paliya2019}.
\citet{paliya2019} estimated the disk luminosity based on multi-wavelength SED modeling for observations, assuming that the optical-UV data are emissions from the disk.
\citet{doi2012} estimated the mechanical luminosity of the outflow by calculating the power of the X-ray cavity from the power of the radio lobe over the kpc scale obtained by Very Large Array (VLA).

If we apply the fitting formula for $L_{\rm mech}/L_{\rm Edd}$,
we require ${\dot m}_{\rm BH} \sim 90$, in order to reproduce the normalized mechanical luminosity of 0.05, 
leading to the normalized isotropic radiation luminosity to be
$L_{\rm rad}^{\rm ISO}/ L_{\rm Edd} \sim (3.3, ~3.1, ~3.1, ~3.1)$ for 
$\theta$ = (10, 20, 30, 60), respectively

Note that the normalized isotropic radiation luminosity is $3.1 - 3.3$ at $\theta=10^\circ - \theta=60^\circ$. 
The luminosity ratio, is approximately, $L_{\rm mech}/L_{\rm rad}^{\rm ISO}\sim 0.02$.
The radiation luminosity is about three times larger than the observed radiation luminosity.

\subsubsection{Why are the simulated radiation luminosities so high?}
\label{sec4.3.5}
We believe that our evaluated mechanical luminosities are more reliable than those in previous numerical simulations, 
since we perform RHD simulations with large enough Keplerian radius ($r_{\rm K} = 10^3 r_{\rm S}$)
in a large simulation box with size of $6 \times 10^3~r_{\rm S}$ to avoid numerical artefacts arising from outflow from the initial torus
(see K21 and references therein).
From the comparison with the observations of NLS1s, however, we find
that the radiation luminosities are about 3 times larger than the observed luminosities in both cases. What is a reason for such a discrepancy?

A possible reason is that our simulations do not consider effects of magnetic fields nor general relativistic effects. 
If we include those effects, we expect that the radiation luminosity could be reduced, since it will be partly transported to the Poynting flux, especially when the black hole spin is high.
In fact, the previous studies based on the general relativistic (GR)-RMHD simulations suggest that 
the Poynting flux can dominate kinetic and radiation energy transport in jets and winds for high-spin cases \citep[e.g.,][]{yang2023, utsumi2022}.
We will investigate the energy transport processes to the Poynting flux in future work.

\section{Concluding remarks}
In this paper, 
we perform two-dimensional axisymmetric RHD simulations to investigate the properties of radiation and outflow 
as functions of the normalized black hole mass $m_{\rm BH}$ and the normalized accretion rate onto the black hole $\dot{m}_{\rm BH}$.
We have the following results, some of which are unexpected before the present study.

\begin{itemize}

  \item 
    We confirm that the $\dot{m}_{\rm BH}$-dependence of the normalized radiation luminosity ($L_{\rm rad}/L_{\rm Edd}$) and mechanical luminosity ($L_{\rm mech}/L_{\rm Edd}$) 
    found for a 10 $M_\odot$ the cases of stellar-mass black holes can apply to the cases of massive black holes.
  
  \item 
    We notice that normalized isotropic mechanical luminosity exhibit a broken power-law distribution.
    For low normalized accretion rates ($\dot{m}_{\rm BH} < 400$), 
    the normalized isotropic mechanical luminosity rapidly grows in proportion to $\dot{m}_{\rm BH}^{2.7}$ as normalized accretion rate increases, 
    while for high normalized accretion rates it increases more slowly ($\propto \dot{m}_{\rm BH}^{0.7}$).
    
  \item 
    The lower $\dot{m}_{\rm BH}-$dependence of the normalized isotropic mechanical luminosity at high $\dot{m}_{\rm BH}$ is due to the rapid decrease in velocity above the break as  $\dot{m}_{\rm BH}$ increases.
    The reason for the rapid decrease in normalized velocity is that the vertically inflated disk causes slower and dense gas to be ejected in the poleward direction.

  \item 
    We find that the $\dot{m}_{\rm BH}-$dependence of the normalized luminosities is insensitive to the black hole mass.
    The normalized mechanical luminosity grows more rapidly than the normalized radiation luminosity with an increase of $\dot{m}_{\rm BH}$,
    even in the case of super-massive black holes.
    Such universalities can appear in the regime, in which electron scattering opacity dominates over absorption opacity.

  \item 
    As future issues we need to solve the general relativistic magnetohydro-dynamics, 
    since then MHD driven outflow and Blandford-Znajek type jet will appear \citep{Blandford1977}.
    In the rapidly spinning ker black hole, 
    the Blandford-Znajek effect causes energy injection through the Poynting flux into the gas near the black hole, 
    which will lead to significant enhancement of the mechanical power of the outflow.
    In the case of super-massive black holes, it has been pointed out that the iron opacity bump,
     which is not considered in our simulations, could affect the disk structure and luminosity variations. 
     This is also an issue for future work.
\end{itemize}

\begin{ack}
  We are also gratefull to an anonymous referee for his/her valuable and constructive comments, which helped us improving our manuscript in a great deal.
  This work was supported in part by JST SPRING Grant No. JPMJSP2110, JSPS Grant-in-Aid for Scientific Research (A) JP21H04488 (KO), and for Scientific Research (C) JP20K04026 (SM), and JP23K03448 (TK).
  This work was also supported by MEXT as “Program for Promoting Researches on the Supercomputer Fugaku” (Structure and Evolution of the Universe Unraveled by Fusion of Simulation and AI; Grant Number JPMXP1020230406; KO, TK), 
  and by the HPCI System Research Project (Project ID: hp230116, hp240054; KO, TK),
  and by Joint Institute for Computational Fundamental Science (JICFuS, KO).
  Numerical computations were in part carried out on
  Cray XC50 at Center for Computational Astrophysics,
  National Astronomical Observatory of Japan. 
\end{ack}


\begin{thebibliography}{}
  \expandafter\ifx\csname natexlab\endcsname\relax\def\natexlab#1{#1}\fi
  
  \bibitem[{Abramowicz {et~al.}(1988)Abramowicz, Czerny, Lasota, \&
    Szuszkiewicz}]{Abrmwiz1988}
  Abramowicz, M., Czerny, B., Lasota, J., \& Szuszkiewicz, E. 1988, \apj, 332, 646
  
  \bibitem[{Bachetti {et~al.}(2014)Bachetti, Harrison, Walton, Grefenstette,
    Chakrabarty, F{\"u}rst, Barret, Beloborodov, Boggs, Christensen,
    {et~al.}}]{Bachetti2014}
  Bachetti, M., Harrison, F., Walton, D.~J., {et~al.} 2014, Nature, 514, 202
  
  \bibitem[{Basko \& Sunyaev(1976)}]{Bsk1976}
  Basko, M., \& Sunyaev, R.~A. 1976, \mnras, 175, 395
  
  \bibitem[{Blandford \& Znajek(1977)}]{Blandford1977}
  Blandford, R.~D., \& Znajek, R.~L. 1977, \mnras, 179, 433
  
  \bibitem[{Botella {et~al.}(2022)Botella, Mineshige, Kitaki, Ohsuga, \&
    Kawashima}]{botella2022}
  Botella, I., Mineshige, S., Kitaki, T., Ohsuga, K., \& Kawashima, T. 2022,
    \pasj, 74, 384
  
  \bibitem[{Cappi {et~al.}(2009)Cappi, Tombesi, Bianchi, Dadina, Giustini,
    Malaguti, Maraschi, Palumbo, Petrucci, Ponti, {et~al.}}]{cappi2009}
  Cappi, M., Tombesi, F., Bianchi, S., {et~al.} 2009, A\&A,
    504, 401
  
  \bibitem[{Carpano {et~al.}(2018)Carpano, Haberl, Maitra, \&
    Vasilopoulos}]{carpano2018}
  Carpano, S., Haberl, F., Maitra, C., \& Vasilopoulos, G. 2018, \mnras, 476, L45
  
  \bibitem[{Doi {et~al.}(2012)Doi, Nagira, Kawakatu, Kino, Nagai, \&
    Asada}]{doi2012}
  Doi, A., Nagira, H., Kawakatu, N., {et~al.} 2012, \apj,
    760, 41
  
  \bibitem[{Done \& Jin(2016)}]{done2016}
  Done, C., \& Jin, C. 2016, \mnras,
    460, 1716
  
  \bibitem[{D’Ammando(2019)}]{DAmmand2019}
  D’Ammando, F. 2019, Galaxies, 7, 87
  
  \bibitem[{Eggum {et~al.}(1988)Eggum, Coroniti, \& Katz}]{eggum1988}
  Eggum, G., Coroniti, F., \& Katz, J. 1988, \apj, 330, 142
  
  \bibitem[{Fabbiano(1989)}]{fabbiano1989}
  Fabbiano, G. 1989, A\&A, 27, 87
  
  \bibitem[{Fujita \& Okuda(1998)}]{fujita1998}
  Fujita, M., \& Okuda, T. 1998, \pasj, 50, 639
  
  \bibitem[{F{\"u}rst {et~al.}(2016)F{\"u}rst, Walton, Harrison, Stern, Barret,
    Brightman, Fabian, Grefenstette, Madsen, Middleton, {et~al.}}]{furst2016}
  F{\"u}rst, F., Walton, D., Harrison, F., {et~al.} 2016, Astrophysical Journal (Letter), 831, L14
  

  \bibitem[{Gladstone {et~al.}(2009)Gladstone, Roberts, \& Done}]{gladstone2009}
  Gladstone, J.~C., Roberts, T.~P., \& Done, C. 2009, \mnras, 397, 1836
  
  \bibitem[{Honma {et~al.}(1991)Honma, Matsumoto, \& Kato}]{honma1991}
  Honma, F., Matsumoto, R., \& Kato, S. 1991, \pasj, 43, 147
  
  \bibitem[{Inoue {et~al.}(2023)Inoue, Ohsuga, Takahashi, \& Asahina}]{inoue2023}
  Inoue, A., Ohsuga, K., Takahashi, H.~R., \& Asahina, Y. 2023, arXiv preprint
    arXiv:2305.12373
  
  \bibitem[{Israel {et~al.}(2017)Israel, Belfiore, Stella, Esposito, Casella,
    De~Luca, Marelli, Papitto, Perri, Puccetti, {et~al.}}]{israel2017}
  Israel, G.~L., Belfiore, A., Stella, L., {et~al.} 2017, Science, 355, 817
  
  \bibitem[{Jiang \& Blaes(2020)}]{jiang2020}
  Jiang, Y.-F., \& Blaes, O. 2020, The Astrophysical Journal, 900, 25
  
  \bibitem[{Jin {et~al.}(2017)Jin, Done, Ward, \& Gardner}]{jin2017}
  Jin, C., Done, C., Ward, M., \& Gardner, E. 2017, \mnras, 471, 706
  
  \bibitem[{Jin {et~al.}(2022)Jin, Done, Ward, Panessa, Liu, \& Liu}]{jin2022}
  Jin, C., Done, C., Ward, M., {et~al.} 2022, \mnras, 512, 5642
  
  \bibitem[{Kaaret {et~al.}(2017)Kaaret, Feng, \& Roberts}]{kaaret2017}
  Kaaret, P., Feng, H., \& Roberts, T.~P. 2017, A\&A, 55, 303
  
  \bibitem[{Kato {et~al.}(2008)Kato, Fukue, \& Mineshige}]{kato2008}
  Kato, S., Fukue, J., \& Mineshige, S. 2008, Black-Hole Accretion
    Disks---Towards a New Paradigm---
  
  \bibitem[{Kawashima {et~al.}(2009)Kawashima, Ohsuga, Mineshige, Heinzeller,
    Takabe, \& Matsumoto}]{kawashima2009}
  Kawashima, T., Ohsuga, K., Mineshige, S., {et~al.} 2009, \pasj, 61, 769
  
  \bibitem[{Kawashima {et~al.}(2012)Kawashima, Ohsuga, Mineshige, Yoshida,
    Heinzeller, \& Matsumoto}]{kawashima2012}
  Kawashima, T., Ohsuga, K., Mineshige, S., {et~al.} 2012, \apj, 752, 18

  \bibitem[{Kawashima {et~al.}(2016)Kawashima, Mineshige, Ohsuga, \&
    Ogawa}]{kawashima2016}
  Kawashima, T., Mineshige, S., Ohsuga, K., \& Ogawa, T. 2016, \pasj, 68, 83
  
  \bibitem[{King {et~al.}(2001)King, Davies, Ward, Fabbiano, \& Elvis}]{king2001}
  King, A.~R., Davies, M.~B., Ward, M., Fabbiano, G., \& Elvis, M. 2001, \apj, 552, L109
  
  \bibitem[{Kitaki {et~al.}(2017)Kitaki, Mineshige, Ohsuga, \&
    Kawashima}]{kitaki2017}
  Kitaki, T., Mineshige, S., Ohsuga, K., \& Kawashima, T. 2017, \pasj, 69, 92
  
  \bibitem[{Kitaki {et~al.}(2018)Kitaki, Mineshige, Ohsuga, \&
    Kawashima}]{kitaki2018}
  Kitaki, T., Mineshige, S., Ohsuga, K., \& Kawashima, T. 2018, \pasj, 70, 108
  
  \bibitem[{Kitaki {et~al.}(2021)Kitaki, Mineshige, Ohsuga, \&
    Kawashima}]{kitaki2021}
    Kitaki, T., Mineshige, S., Ohsuga, K., \& Kawashima, T. 2021, \pasj, 73, 450 (K21)
  
  \bibitem[{Landt {et~al.}(2017)Landt, Ward, Balokovi{\'c}, Kynoch,
    Storchi-Bergmann, Boisson, Done, Schimoia, \& Stern}]{landt2017}
  Landt, H., Ward, M.~J., Balokovi{\'c}, M., {et~al.} 2017, \mnras, 464, 2565
  
  \bibitem[{Levermore \& Pomraning(1981)}]{levermore1981}
  Levermore, C., \& Pomraning, G. 1981, \apj, 248, 321
  
  \bibitem[{Lister(2016)}]{lister2016}
  Lister, M. 2016, AJ, 152, 12
  
  \bibitem[{Long {et~al.}(1996)Long, Charles, Blair, \& Gordon}]{long1996}
  Long, K.~S., Charles, P.~A., Blair, W.~P., \& Gordon, S.~M. 1996, \apj, 466, 750
  
  \bibitem[{Makishima {et~al.}(2000)Makishima, Kubota, Mizuno, Ohnishi, Tashiro,
    Aruga, Asai, Dotani, Mitsuda, Ueda, {et~al.}}]{makishima2000}
  Makishima, K., Kubota, A., Mizuno, T., {et~al.} 2000, \apj, 535, 632

  \bibitem[{McKinney {et~al.}(2014)McKinney, Tchekhovskoy, Sadowski, \&
    Narayan}]{mckinney2014}
  McKinney, J.~C., Tchekhovskoy, A., Sadowski, A., \& Narayan, R. 2014, \mnras, 441, 3177
  
  \bibitem[{McKinney {et~al.}(2015)McKinney, Dai, \& Avara}]{mckinney2015}
  McKinney, J.~C., Dai, L., \& Avara, M.~J. 2015, \mnras, 454, L6

  
  \bibitem[{Middleton {et~al.}(2015)Middleton, Heil, Pintore, Walton, \&
    Roberts}]{middleton2015}
  Middleton, M.~J., Heil, L., Pintore, F., Walton, D.~J., \& Roberts, T.~P. 2015,
    \mnras, 447, 3243
  
  \bibitem[{Miller {et~al.}(2004)Miller, Fabian, \& Miller}]{miller2004}
  Miller, J.~M., Fabian, A., \& Miller, M. 2004, \apj, 614, L117
  
  \bibitem[{Mineshige {et~al.}(2000)Mineshige, Kawaguchi, Takeuchi, \&
    Hayashida}]{mineshige2000}
  Mineshige, S., Kawaguchi, T., Takeuchi, M., \& Hayashida, K. 2000, \pasj, 52, 499
  
  \bibitem[{Miyawaki {et~al.}(2009)Miyawaki, Makishima, Yamada, Gandhi, Mizuno,
    Kubota, Tsuru, \& Matsumoto}]{miyawaki2009}
  Miyawaki, R., Makishima, K., Yamada, S., {et~al.} 2009, \pasj, 61, S263
  
  \bibitem[{Motch {et~al.}(2014)Motch, Pakull, Soria, Gris{\'e}, \&
    Pietrzy{\'n}ski}]{motch2014}
  Motch, C., Pakull, M., Soria, R., Gris{\'e}, F., \& Pietrzy{\'n}ski, G. 2014,
    Nature, 514, 198
  
  \bibitem[{Mushtukov {et~al.}(2018)Mushtukov, Verhagen, Tsygankov, van~der Klis,
    Lutovinov, \& Larchenkova}]{mushtukov2018}
  Mushtukov, A.~A., Verhagen, P.~A., Tsygankov, S.~S., {et~al.} 2018, \mnras, 474, 5425
  
  \bibitem[{Nardini {et~al.}(2015)Nardini, Reeves, Gofford, Harrison, Risaliti,
    Braito, Costa, Matzeu, Walton, Behar, {et~al.}}]{nardini2015}
  Nardini, E., Reeves, J., Gofford, J., {et~al.} 2015, Science, 347, 860
  
  \bibitem[{Ogawa {et~al.}(2017)Ogawa, Mineshige, Kawashima, Ohsuga, \&
    Hashizume}]{ogawa2017}
  Ogawa, T., Mineshige, S., Kawashima, T., Ohsuga, K., \& Hashizume, K. 2017,
    \pasj, 69, 33
  
  
  \bibitem[{Ohsuga {et~al.}(2005)Ohsuga, Mori, Nakamoto, \&
    Mineshige}]{ohsuga2005}
  Ohsuga, K., Mori, M., Nakamoto, T., \& Mineshige, S. 2005, \apj, 628, 368

  \bibitem[{Ohsuga(2006)}]{ohsuga2006}
  Ohsuga, K. 2006, \apj, 640, 923
  
  \bibitem[{Ohsuga(2007)}]{ohsuga2007}
  Ohsuga, K. 2007, \pasj, 59, 1033

  
  \bibitem[{Paczynsky \& Wiita(1980)}]{paczynsky1980}
  Paczynsky, B., \& Wiita, P.~J. 1980, A\&A, 88, 23
  
  \bibitem[{Paliya {et~al.}(2019)Paliya, Parker, Jiang, Fabian, Brenneman,
    Ajello, \& Hartmann}]{paliya2019}
  Paliya, V.~S., Parker, M., Jiang, J., {et~al.} 2019, \apj,
    872, 169
  
  \bibitem[{Reeves {et~al.}(2003)Reeves, O’Brien, \& Ward}]{reeves2003}
  Reeves, J., O’Brien, P.~T., \& Ward, M. 2003, \apj, 593,
    L65
  
  \bibitem[{Rybicki \& Lightman(1991)}]{rybicki1991}
  Rybicki, G.~B., \& Lightman, A.~P. 1991, Radiative processes in astrophysics
  
  \bibitem[{S{\c{a}}dowski {et~al.}(2014)S{\c{a}}dowski, Narayan, McKinney, \&
    Tchekhovskoy}]{sadowski2014}
  S{\c{a}}dowski, A., Narayan, R., McKinney, J.~C., \& Tchekhovskoy, A. 2014,
    \mnras, 439, 503
  
  \bibitem[{S{\c{a}}dowski {et~al.}(2015)S{\c{a}}dowski, Narayan, Tchekhovskoy,
    Abarca, Zhu, \& McKinney}]{sadowski2015}
  S{\c{a}}dowski, A., Narayan, R., Tchekhovskoy, A., {et~al.} 2015, \mnras, 447, 49

  \bibitem[{S{\c{a}}dowski \& Narayan(2016)}]{sadowski2016}
  S{\c{a}}dowski, A., \& Narayan, R. 2016, \mnras, 456, 3929
  
  \bibitem[{Shakura \& Sunyaev(1973)}]{shakura1973}
  Shakura, N.~I., \& Sunyaev, R.~A. 1973, A\&A, 24, 337
  
  \bibitem[{Soria(2007)}]{soria2007}
  Soria, R. 2007, Ap\&SS, 311, 213
  
  \bibitem[{Strohmayer \& Mushotzky(2009)}]{strohmayer2009}
  Strohmayer, T.~E., \& Mushotzky, R.~F. 2009, \apj, 703, 1386
  
  \bibitem[{Sutton {et~al.}(2013)Sutton, Roberts, \& Middleton}]{sutton2013}
  Sutton, A.~D., Roberts, T.~P., \& Middleton, M.~J. 2013, \mnras, 435, 1758
  
  \bibitem[{Takahashi {et~al.}(2010)Takahashi, Hayashida, \&
    Anabuki}]{takahashi2010}
  Takahashi, H., Hayashida, K., \& Anabuki, N. 2010, \pasj, 62, 1483
  
  \bibitem[{Takahashi {et~al.}(2016)Takahashi, Ohsuga, Kawashima, \&
    Sekiguchi}]{takahashi2016}
  Takahashi, H.~R., Ohsuga, K., Kawashima, T., \& Sekiguchi, Y. 2016, \apj, 826, 23
  
  \bibitem[{Takeo {et~al.}(2018)Takeo, Inayoshi, Ohsuga, Takahashi, \&
    Mineshige}]{takeo2018}
  Takeo, E., Inayoshi, K., Ohsuga, K., Takahashi, H.~R., \& Mineshige, S. 2018,
    \mnras, 476, 673
  
  \bibitem[{Tombesi {et~al.}(2010)Tombesi, Cappi, Reeves, Palumbo, Yaqoob,
    Braito, \& Dadina}]{tombesi2010}
  Tombesi, F., Cappi, M., Reeves, J., {et~al.} 2010, A\&A, 521, A57
  
  \bibitem[{Tombesi {et~al.}(2014)Tombesi, Tazaki, Mushotzky, Ueda, Cappi,
    Gofford, Reeves, \& Guainazzi}]{tombesi2014}
  Tombesi, F., Tazaki, F., Mushotzky, R., {et~al.} 2014, \mnras, 443, 2154
  
  \bibitem[{Tombesi {et~al.}(2015)Tombesi, Mel{\'e}ndez, Veilleux, Reeves,
    Gonz{\'a}lez-Alfonso, \& Reynolds}]{tombesi2015}
  Tombesi, F., Mel{\'e}ndez, M., Veilleux, S., {et~al.} 2015, Nature, 519, 436

  \bibitem[{Turner(2004)}]{turner2004}
  Turner, N. 2004, \apj, 605, L45
  
  \bibitem[{Turner \& Stone(2001)}]{turner2001}
  Turner, N., \& Stone, J. 2001, \apjs, 135, 95
  
  \bibitem[{Utsumi {et~al.}(2022)Utsumi, Ohsuga, Takahashi, \&
    Asahina}]{utsumi2022}
  Utsumi, A., Ohsuga, K., Takahashi, H.~R., \& Asahina, Y. 2022, \apj, 935, 26
  
  \bibitem[{Wang \& Zhou(1999)}]{wang1999}
  Wang, J.-M., \& Zhou, Y.-Y. 1999, \apj, 516, 420
  
  \bibitem[{Watarai {et~al.}(2001)Watarai, Mizuno, \& Mineshige}]{watarai2001}
  Watarai, K.-y., Mizuno, T., \& Mineshige, S. 2001, \apj, 549, L77
  
  \bibitem[{Xu {et~al.}(2021)Xu, Pinto, Bianchi, Kosec, Parker, Walton, Fabian,
    Guainazzi, Barret, \& Cusumano}]{xu2021}
  Xu, Y., Pinto, C., Bianchi, S., {et~al.} 2021, \mnras, 508, 6049
  
  \bibitem[{Yang {et~al.}(2023)Yang, Yuan, Kwan, \& Dai}]{yang2023}
  Yang, H., Yuan, F., Kwan, T., \& Dai, L. 2023, \mnras, 523, 208
  
  \bibitem[{Yoshioka {et~al.}(2022)Yoshioka, Mineshige, Ohsuga, Kawashima, \&
    Kitaki}]{yoshioka2022}
  Yoshioka, S., Mineshige, S., Ohsuga, K., Kawashima, T., \& Kitaki, T. 2022,
    \pasj, 74, 1378 (Paper I)
  
  \bibitem[{Yuan {et~al.}(2008)Yuan, Zhou, Komossa, Dong, Wang, Lu, \&
    Bai}]{yuan2008}
  Yuan, W., Zhou, H., Komossa, S., {et~al.} 2008, \apj, 685, 801
  
  \bibitem[{Zhou {et~al.}(2006)Zhou, Wang, Yuan, Lu, Dong, Wang, \&
    Lu}]{zhou2006}
  Zhou, H., Wang, T., Yuan, W., {et~al.} 2006, \apjs, 166, 128
\end{thebibliography}

\clearpage
%
\appendix

\section{Light curves}

\begin{figure}[]
  \begin{center}
    \includegraphics[width=80mm]{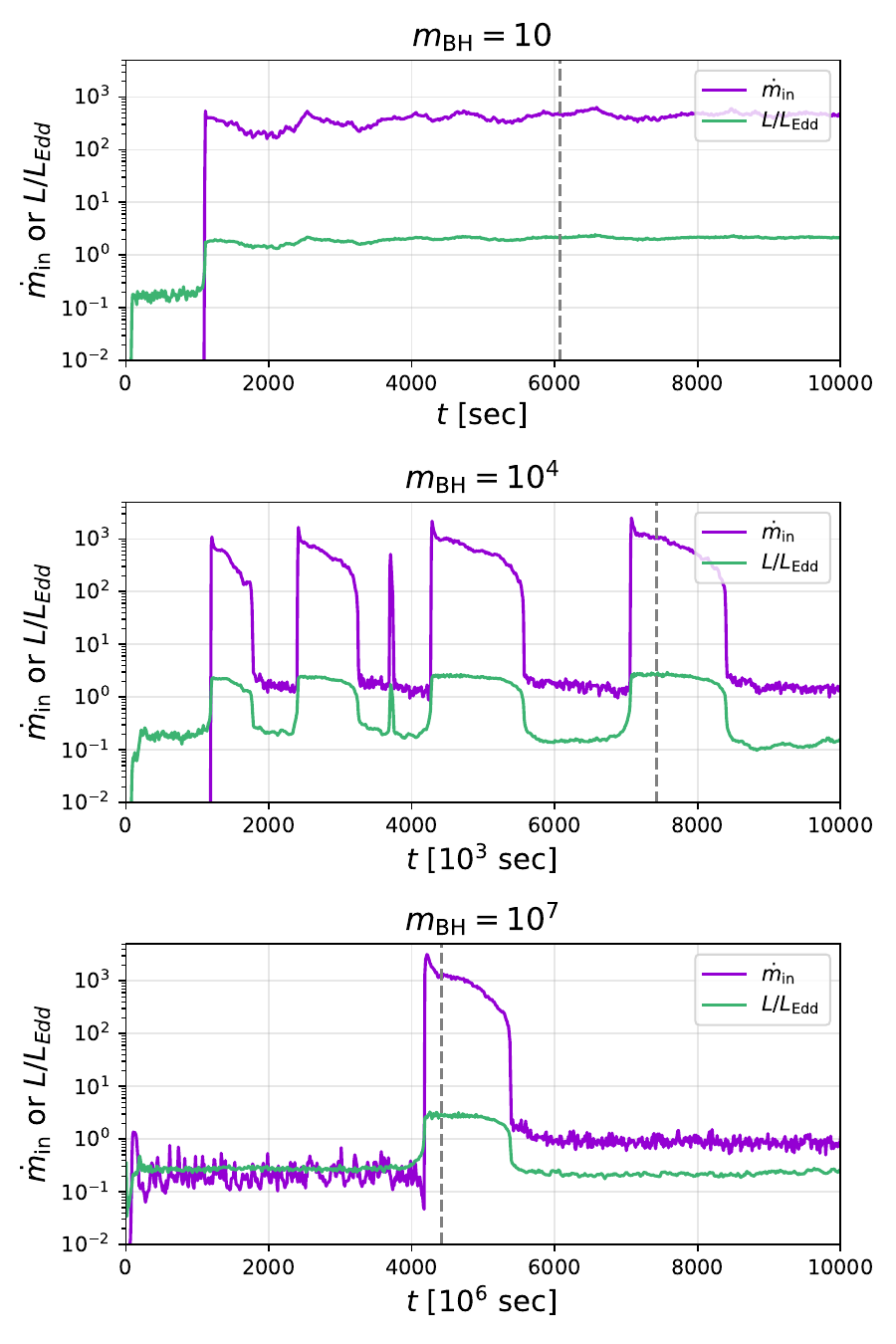}
  \end{center}
  \caption{
    Time evolution of the mass input rates and outgoing radiation luminosities for $m_{\rm BH} = 10^1$ (top), $10^4$ (middle) and $10^7$ (bottom), respectively.
    The purple and green lines represent the normalized accretion rate at $r=r_{\rm in}=20~r_{\rm S}$ and the radiation luminosity in the laboratory frame, respectively.
    The grey dashed line represents the start time of the second-step simulation for each model.
  }
  \label{figa1}
\end{figure}

Figure \ref{figa1} show time evolution of the normalized accretion rate $\dot{m}_{\rm  in}$ (purple line) and 
normlized luminosity in the laboratory frame $L/L_{\rm Edd}$ (green line) for $m_{\rm BH} = 10^1,~10^4,~10^7$ in the first-step simulations. 
We calculate the normalized accretion rate at $r = 20~r_{\rm S}$ and the normalized luminosity 
at $r = r_{\rm out}$ by 
\begin{eqnarray}
  \dot{m}_{\rm in} & \equiv & 4\pi \frac{c^2}{L_{\rm Edd}}\int_{0}^{\pi/2} d\theta\sin\theta \times \left(20~r_{\rm S}\right)^{2}\nonumber\\
  &&~~~~~~~~~\times \rho(20~r_{\rm S},\theta)~|{\rm min}\left\{v_{r}(20~r_{\rm S},\theta),0\right\}|,
\end{eqnarray}
\begin{eqnarray}
  L& \equiv & 4\pi\int_{0}^{\pi/2} d\theta\sin\theta \times r_{\rm out}^2~ {\rm max}\left\{F_{\rm lab}^r,~0\right\}.
\end{eqnarray}
We should note that the definitions and the absolute values of $\dot{m}_{\rm in}$ and $L/L_{\rm Edd}$ are slightly 
different from $\dot{m}_{\rm BH}$ and $L_{\rm rad}/L_{\rm Edd}$ used in the text. 
Figure \ref{figa1} shows that the high-low transitions between the super-Eddington and sub-Eddington states occur at constant intervals.
This is the same type of limit cycle oscillation \citep[e.g.,][]{Abrmwiz1988, honma1991, ohsuga2006, ohsuga2007}.
It is due to the viscous instability that occurs when radiation pressure is dominant.
For the transitions described above, we started a second-step simulation at the upward transition (grey dashed line).
The start times of the second-step simulations are $6070$, $7430\times10^3$, and $4420\times10^6$ sec with $m_{\rm BH}=10$, $10^4$, and $10^7$ respectively.

\end{document}